\documentclass[floatfix, longbibliography,aps,pre,superscriptaddress,showpacs,twocolumn]{revtex4-1}

\usepackage{geometry}
\usepackage{float}

\usepackage{amsmath}
\usepackage{amssymb}
\usepackage{graphicx}

\usepackage[]{algorithm2e}
\usepackage{booktabs}
\usepackage{setspace}
\usepackage[toc,page]{appendix}
\usepackage[justification=justified,format=plain]{caption}
\usepackage{xcolor}
\usepackage{xr}
\usepackage{amssymb}

\usepackage{tabularx}
\usepackage{hyperref}

\begin{document}

	\global\long\def\d{\mathrm{d}}
	\global\long\def\v#1{\mathrm{\mathbf{#1}}}
	\global\long\def\var{\mathrm{var}}
	\global\long\def\unit#1{\,\mathrm{#1}}
	\global\long\def\ParBS{$\mathrm{ParB}S$\,}
	\global\long\def\dParBS{$\Delta\mathrm{ParB}/S$\,}
	\newcommand{\kMotor}{k_\mathrm{motor}}
	\newcommand{\vMotor}{v_\mathrm{motor}}
	\newcommand{\kMotorRed}{\tilde{k}}
	
	\title{Bacterial chromosome organization by collective dynamics of SMC condensins}
	
	\author{Christiaan A. Miermans} 
	\affiliation{Arnold-Sommerfeld-Center for Theoretical Physics and Center for
		NanoScience, Ludwig-Maximilians-Universit\"at M\"unchen,
		D-80333 M\"unchen, Germany.}
	\author{Chase P. Broedersz}
	\email{C.broedersz@lmu.de}
	\affiliation{Arnold-Sommerfeld-Center for Theoretical Physics and Center for
		NanoScience, Ludwig-Maximilians-Universit\"at M\"unchen,
		D-80333 M\"unchen, Germany.}

	\date{\today}
	
	\begin{abstract}
	A prominent organizational feature of bacterial chromosomes was revealed by Hi-C experiments, indicating anomalously high contacts between the left and right chromosomal arms. These long-range contacts have been attributed to various nucleoid-associated proteins, including the ATPase SMC condensin. Although the molecular structure of these ATPases has been mapped in detail, it still remains unclear by which physical mechanisms they collectively generate long-range chromosomal contacts. Here, we develop a computational model that captures the subtle interplay between molecular-scale activity  of slip-links and large-scale chromosome organization. We first consider a scenario in which the ATPase activity of slip-links regulates their DNA-recruitment near the origin of replication, while the slip-link dynamics is assumed to be diffusive. We find that such diffusive slip-links can collectively organize the entire chromosome into a state with aligned arms, but not within physiological constraints. However, slip-links that include motor activity are far more effective at organizing the entire chromosome over all length-scales. The persistence of motor slip-links  at physiological densities can  generate large, nested loops and drive them into the bulk of the DNA. Finally, our model with motor slip-links can quantitatively account for the rapid arm--arm  alignment  of chromosomal arms observed \textit{in vivo}.
\end{abstract}

	\maketitle
	\section{Introduction}
	The bacterial chromosome is highly structured over a wide range of length-scales~\cite{Wang2014, Gruber2014a, Badrinarayanan2015, Kleckner2014}.
	 Indeed, microscopy experiments have revealed a remarkable degree of sub-cellular organization of the chromosome across many species, including \textit{Bacillus subtilis}, \textit{Caulobacter crescentus} and \textit{Escherichia coli}~\cite{Viollier2004, Marbouty2015, Wiggins2010, Wu2018,Sullivan}.
	 The structural organization of chromosomes was further exposed by recent Hi-C experiments measuring the contact probability between pairs of loci on the chromosome~\cite{Le2013,Wang2015a, Wang2017a, Marbouty2015, Tran2017}.
	On small length-scales, the chromosome appears to be organized in so-called Chromosome Interaction Domains: genomic domains with above-average contact probabilities between pairs of loci. On large length-scales, the most prominent feature is the emergence of a cross-diagonal in Hi-C maps spanning the whole length of the genome, indicating anomalously high contact probabilities between opposing pairs of DNA-loci positioned on the left and right chromosomal arms. Such a cross-diagonal is observed in  \textit{B.~subtilis} and  \textit{C.~crescentus}~\cite{Wang2015a,Wang2017a,Le2013, Umbarger2011}, but not in  \textit{E.~coli}~\cite{Lioy2018}.
	
	The measured cross-diagonal indicates a robust juxtaposition of the left and right arms of the chromosome~\cite{Le2013,Wang2017a,Wang2015a, Umbarger2011}, an organizational feature that is important for faithful chromosome segregation~\cite{Gruber2009,  Marbouty2015, Gruber2014, Sullivan, Wang2014b}. This juxtaposed organization is largely controlled by the highly conserved ATPase SMC condensin~\cite{Wang2015a,Wang2017a, Marbouty2015}. While much is known about condensin at the molecular level~\cite{Soh2015,Burmann2013,Hirano2006a, Diebold-Durand2017,  Eeftens2016, Kamada2017}, it is unclear how small numbers of condensins (3--30 per chromosome~\cite{Wilhelm2015})
	are capable of collectively organizing the chromosome over such a range of length-scales. Thus, the physical principles underlying the  juxtaposed organization of the chromosome remain elusive.
	
	The functional capability of SMC condensins to organize the chromosome into a juxtaposed state crucially depends on two factors:\,(i)\,the presence of a specific loading site on the chromosome~\cite{Wang2015a, Minnen2016}, and (ii)\,the ability of condensin to bind and hydrolyze ATP~\cite{Minnen2016, Wilhelm2015}.
	In \textit{B. subtilis}, the loading site is established  by a large nucleoprotein complex composed of ParB proteins bound around  \textit{parS} close to the origin of replication (\textit{ori})~\cite{Broedersz2014, Mohl1997, surtees2003plasmid, Livny2007}.
	Condensins are recruited to this \ParBS region, and from there propagate deep into the bulk of the DNA polymer~\cite{Minnen2016, Wilhelm2015}. Removing the loading site results in a  loss of the cross-diagonal in Hi-C maps, and adding additional loading sites disrupts the cross-diagonal~\cite{Wang2015a, Wang2017a}. 
	Moreover, the translocation of condensins away from the loading site depends on their ATPase activity; condensin mutants that cannot bind ATP only weakly associate with DNA, and mutants that do not hydrolyze ATP do not efficiently propagate away from the \ParBS loading site (SI~\ref{sec:empirical-metrics} and~\cite{Minnen2016, Wilhelm2015}). With a ring-like topology of $25-50\unit{nm}$ in diameter, the  condensin ring is large enough to trap a DNA loop by threading a DNA duplex through it~\cite{Eeftens2016, Burmann2017, Soh2015,Wilhelm2015,Hirano2006a,Diebold-Durand2017, Burmann2017, Burmann2013}.  It has been proposed that the possibility of condensins to trap DNA-loops would enable them to align the chromosomal arms by progressively extruding DNA-loops from the origin to the terminus region~\cite{Wang2015a, Wang2017a, Marbouty2015}.

	Important clues on the role of ATPase activity in SMC condensin come from \textit{in vitro} single-molecule  experiments. These experiments revealed that  \textit{Saccharomyces cerevisiae}  condensin is a molecular motor, and performs active translocation over DNA duplexes~\cite{Terekawa2017}. Kymographs showed that the direction of movement of yeast condensin is  \textit{a priori} random, and switches direction after a typical time-scale; in other words, yeast  condensin performs persistent random motion over DNA.  In fact, recent single-molecule experiments have revealed that such yeast condensin performs active loop extrusion~\cite{Ganji2018}.
	Similar experiments have not yet observed such motor activity for bacterial condensin~\cite{Kim2016}. Nevertheless, it has been widely speculated that the ATPase activity of bacterial condensin is also directed towards motor activity~\cite{Eeftens2017, Burmann2017, Wilhelm2015, Diebold-Durand2017}.
	In this picture, condensin would actively extrude DNA-loops, possibly by feeding DNA duplexes through its ring-like structure~\cite{Sanborn, Fudenberg2016, Brackley2016, Alipour2012}. In contrast, other models have been proposed in which the ATPase activity of condensin is directed towards regulating its association with DNA~\cite{Wilhelm2015, Minnen2016, Brackley2016, Brackley2018}. Thus, it remains an open question whether bacterial condensin also acts as a loop extruding enzyme, or whether the ATPase activity primarily regulates the DNA-recruitment of condensins.

\begin{figure}[t!]

		\includegraphics[width=\linewidth]{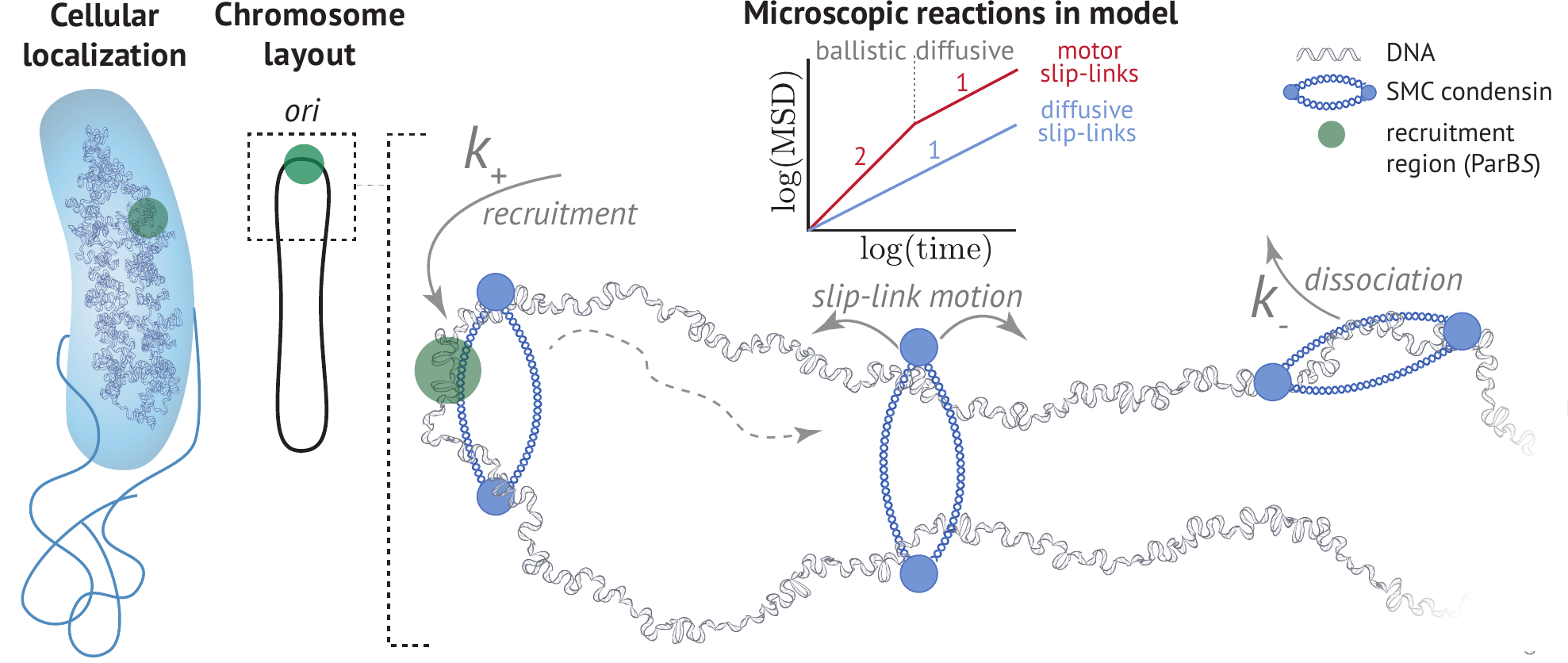}
		\caption{ {\small Schematic of  slip-link model for SMC condensin--DNA interactions.  \textbf{Left}: The bacterial chromosome has a circular topology. SMC condensins are loaded onto the DNA by the ParB\textit{S} complex located near the origin of replication (\textit{ori}, green). \textbf{Right}: Microscopic reactions in our model with associated rate constants. Condensins are represented as slip-links, which either perform diffusive or more persistent, motorized movement (inset). Our computational method simulates the stochastic interplay between DNA dynamics and slip-link positioning.}
			\label{fig:Microscopic-reactions}
		}
\end{figure}

	To elucidate the role of condensin ATPase activity on bacterial chromosome organization, we develop two minimal models for condensin--DNA interactions, where we describe SMC condensin as a slip-link that non-topologically traps a DNA-loop (Fig.~\ref{fig:Microscopic-reactions}). In these models, we analyze the complex interplay between the molecular-scale  dynamics of SMC condensins and large-scale chromosome organization.  In the most basic model, condensin activity is assumed to be directed towards regulating its DNA-recruitment, while the dynamics of condensin slip-links on the DNA is diffusive. 
	Although the motion of individual slip-links on the DNA is purely diffusive, their active recruitment to DNA results in non-equilibrium collective motion of slip-links. Interestingly, we find that such  \textit{diffusive slip-links} can organize the chromosome into a juxtaposed state, but not within physiological constraints.  Next, we expand the model to include  \textit{motor slip-links} that perform persistent random motion on DNA, as observed for yeast condensin\textit{in vitro}~\cite{Terekawa2017}. We find that these motor slip-links are much more effective in organizing the entire chromosome. In particular, our motor slip-link model requires at least 2--3 orders of magnitude fewer condensins to organize the chromosome than in the diffusive slip-link model.
	In addition, the development of the juxtaposed state exhibits sub-diffusive dynamics in the diffusive slip-link model, in contradiction with the rapid re-organization observed  \textit{in vivo}~\cite{Wang2017a}. We show that such a fast re-organization of the chromosome can be achieved by motor activity in the form of active loop extrusion.  More generally, we provide a quantitative model to address key questions in bacterial chromosome organization, such as the role of an exclusive loading site, cell confinement, motor activity and interaction of SMC condensin with other DNA-bound factors.

	\section{Model}

	Our model of condensin--DNA interactions (Fig.~\ref{fig:Microscopic-reactions}) contains two  ingredients: (i)\,a circular DNA polymer, and (ii)\,multiple SMC condensins that interact with the DNA. We employ a  lattice polymer with a lattice constant set by the persistence length of DNA (SI~\ref{sec:Explanation-of-KMC}). Condensins are modeled as  slip-links: elastic rings that trap a DNA-loop by encircling two DNA duplexes. Importantly, we consider both diffusive and motor slip-links. Diffusive slip-links move randomly (stepping rate $k_0$) over the DNA  (Fig.~\ref{fig:Microscopic-reactions}, inset); motor slip-links perform persistent random motion (translocation rate $k_\mathrm{motor}$) with a persistence time-scale $\tau_\mathrm{switch} = k_\mathrm{switch}^{-1}$. These persistent dynamics enable motor slip-links to actively extrude DNA loops. 
	
	To simulate the dynamics of both the DNA polymer and the slip-links, we developed a Kinetic Monte-Carlo (KMC) algorithm (SI~\ref{sec:Explanation-of-KMC}). Our KMC algorithm simulates the  Rouse dynamics of DNA~\cite{Weber2010, Doi1986}, the associated stochastic motion of slip-links on the DNA, as well as the microscopic reactions in which slip-links bind to (rate $k_+$) or unbind from (rate $k_-$) the DNA. For simplicity, we assume instantaneous slip-link binding $k_+ \rightarrow \infty$, justified by the relatively fast cytosolic diffusion of condensin~\cite{Stigler2016a}.
	Additionally, we fix a  maximum number of slip-links $N_p$ that can bind to the DNA.
	
	
	\section*{Results}
	\subsection{Diffusive slip-links with a specific loading site can organize the chromosome}
\begin{figure}[t!]
			\includegraphics[width=0.9\linewidth]{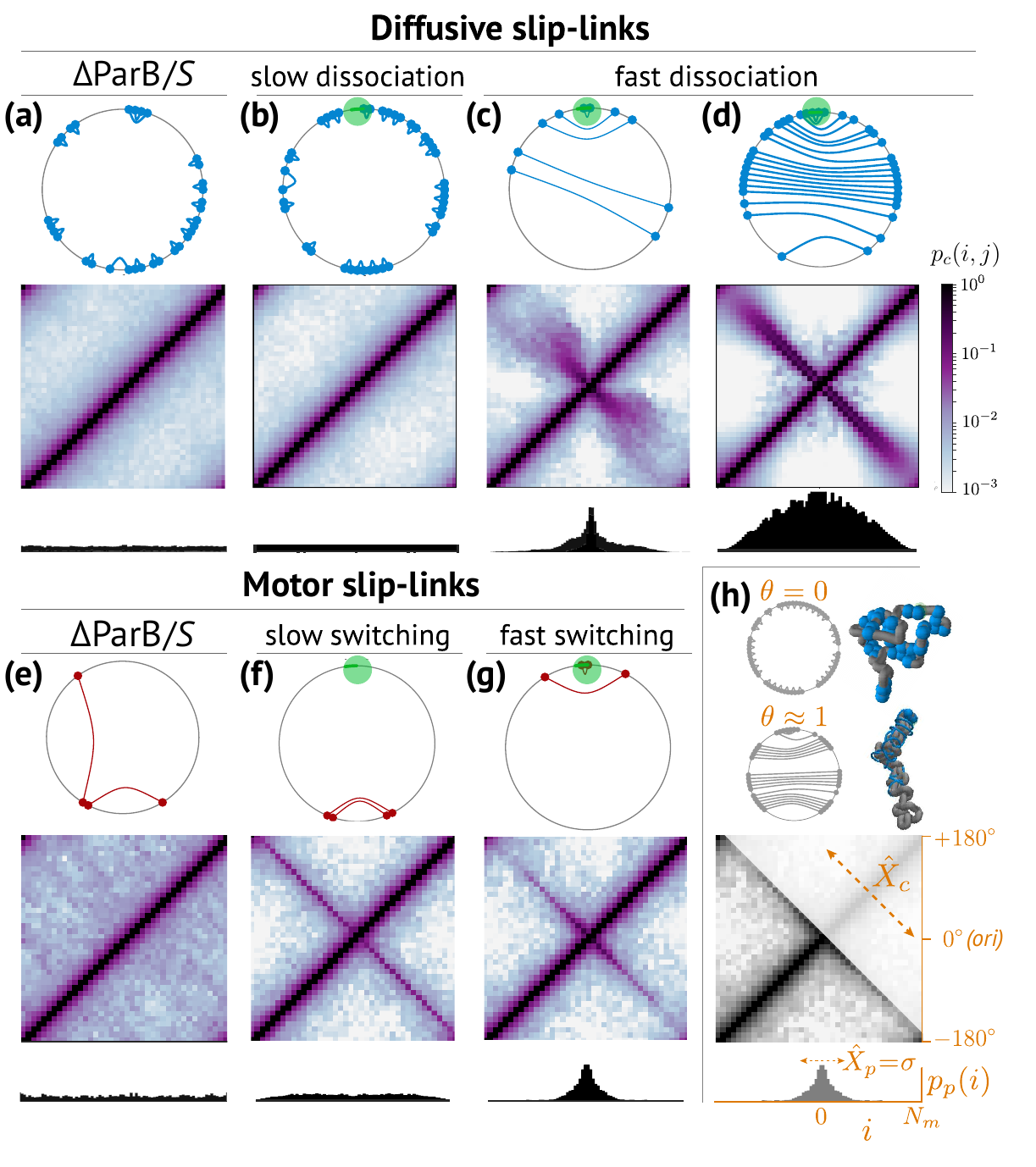}
			\caption{
				{\small Steady-state organization of a circular chromosome by diffusive and motor slip-links. For all sub-figures (a)--(h): top panels show a loop diagram that illustrates the loop structure induced by slip-links, middle panels the contact map $p_c(i,j)$, and bottom panels the binding profile $p_p(i)$ of slips-links along the DNA. \textbf{Diffusive slip-links, (a)--(d)}: (a) lacks a specific loading site (\dParBS), (b) $k_- = 10^{-6}k_0$, (c)--(d) $k_- = k_0$ with $N_p=5$ in (c) and $N_p=25$ in (d). \textbf{Motor slip-links, (e)--(g)} (e) lacks a specific loading site (\dParBS), (f) $k_{\rm switch}=10^{-6}k_\mathrm{motor}$, (g) $k_{\rm switch}=10^{-1} k_\mathrm{motor}$. All simulations were performed with a circular polymer polymer of length $N_m=80$. Maximum number of polymer-bound slip-links from (a)--(g) are respectively $N_p = 16,20,5,25,2,2,2$. \textbf{Legend, (h)}: We use the fraction of nested loops ($\theta$, see SI~\ref{sec:measurement-of-metrics-from-simulation-data}), cross-diagonal length ($\hat X_c$, see SI~\ref{sec:measurement-of-metrics-from-simulation-data}) and the slip-link propagation distance ($\hat X_p=\sqrt  {\var \ p_p(i)  }$) to quantify  polymer organization. The slip-link loading site (when present) is positioned in the center of the contact map, so that the axes run from
				$-\tfrac{1}{2}N_m\ldots+\tfrac{1}{2}N_m$. }
				\label{fig:overview-stationary-results} 
			}
	\end{figure}

	In the simplest implementation of our model, there is no specific loading site: diffusive slip-links can bind and unbind  anywhere on the DNA (Fig.~\ref{fig:overview-stationary-results}a, \dParBS). This assumption results in a homogeneous binding probability $p_p(i)$ (Fig.~\ref{fig:overview-stationary-results}a, bottom).  Importantly, in the \dParBS scenario, all  microscopic reactions are fully reversible, implying that the system relaxes into thermodynamic equilibrium. In  equilibrium, small loops are strongly favored, owing to the increasing entropic cost of loop formation with larger loop size. This tendency to form small loops is reflected in the loop diagrams (Fig.~\ref{fig:overview-stationary-results}a, top).  Indeed, the loop sizes trapped by the slip-links are consistent with the equilibrium loop-size distribution (SI~\ref{sec:diff-slip-links-contact-distribution}). Furthermore,  in this scenario the contact maps are structureless and only exhibit a single main diagonal (Fig.~\ref{fig:overview-stationary-results}a, middle), as  for a random polymer. These results demonstrate that our KMC model with reversible microscopic reactions evolves towards thermodynamic equilibrium. Thus, although the slip-links can easily bind over the full extent of the polymer, they do not organize the chromosome.
	We next investigate how the presence of a slip-link loading site~\cite{Minnen2016, Wilhelm2015, Wang2015a, Wang2017a, Gruber2009} impacts  steady-state chromosome organization.  Note, while recruitment of slip-links  in our model is exclusive to  \textit{ori}, unbinding can occur anywhere on the DNA. This implies that the reactions involving slip-link binding/unbinding are partially irreversible. Hence, detailed balance is broken~\cite{Gnesotto2018}, and the system may no longer  evolve towards thermodynamic equilibrium. Nevertheless, for slow dissociation kinetics (small $k_-$), we observe an unstructured contact map (Fig.~\ref{fig:overview-stationary-results}b; SI~\ref{sec:diff-slip-links-contact-distribution}),  similar to the equilibrium system lacking a specific loading site (Fig.~\ref{fig:overview-stationary-results}a; SI~\ref{sec:diff-slip-links-contact-distribution}). Moreover, the loop diagrams again show that the slip-links mostly encircle small loops (Fig.~\ref{fig:overview-stationary-results}b, top). In sum, we see that, although detailed balance is broken on the level of slip-link binding/unbinding, the diffusive slip-links with slow dissociation kinetics do not appear to organize the DNA polymer.
	
	Interestingly, increasing the dissociation kinetics of slip-links results in dramatically different contact map. We observe the emergence of a prominent cross-diagonal, which either disperses away from the loading site (Fig.~\ref{fig:overview-stationary-results}c) or remains clearly resolved over the whole polymer (Fig.~\ref{fig:overview-stationary-results}d), depending on the density of slip-links.  Interestingly, the loop diagrams for these systems exhibit a topology distinct from the equilibrium configuration; slip-links trap DNA-loops in a cooperative, nested manner. Movies of the loop diagrams and contact maps clearly demonstrate that these nested loops propagate away from the loading site, dynamically driving a juxtaposition of the two polymer arms (SI movies\,3a--b; SI~\ref{section:relaxatin-juxtaposed-organization}). This dynamical arm--arm alignment is a distinct out-of-equilibrium phenomenon, and thus requires the exclusive binding of slip-links to the loading site. Thus, our observations show that diffusive slip-links with fast dissociation kinetics in conjunction with a loading site can, in principle, generate a non-equilibrium polymer organization similar to that found in living cells~\cite{Wang2015a,Wang2017a, Marbouty2015, Le2013}.
	
\begin{figure}
		\includegraphics[width=\linewidth]{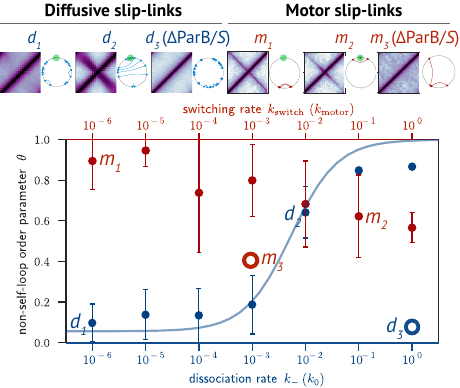}
		\caption{{\small Loop network topology controls polymer organization. The fraction of nested loops $\theta$ of diffusive slip-links (blue)  is shown as a function of the dissociation rate $k_-$  (in units of the slip-link diffusion rate $k_0$), and for motor slip-links (red) as a function of the switching rate $k_\mathrm{switch}$ (in units of the slip-link translocation rate $k_\mathrm{motor}$). Error bars represent the standard deviation $\sqrt{\var{\theta}}$. A  mean-field model  yields an estimate for $\theta$  of diffusive slip-links  (blue curve). Representative loop diagrams and contact maps are indicated for different parameter choices, labeled $d_1$--$d_3$ for diffusive links and $m_1$--$m_3$ for motor slip-links. Also shown are $\theta$ for diffusive ($d_3$) and motor ($m_3$) slip-links with non-specific slip-link binding (\dParBS).  Polymer lengths $N_m$ and slip-link numbers $N_p$ are $N_m=40, N_p=10$ (diffusive slip-links) and $N_m=80,N_p=2$ (motor slip-links).}
			\label{fig:stationary-phenomology}
		}
	\end{figure}

	To investigate the role of slip-link kinetics on loop topology more quantitatively, we employ a metric that captures  essential topological differences between the loop network of random and juxtaposed polymers (compare Fig.~\ref{fig:overview-stationary-results}b to Figs.~\ref{fig:overview-stationary-results}c--d). To this end, we define the order parameter $\theta  \in [0,1] $ as the fraction of nested loops (Fig.~\ref{fig:overview-stationary-results}h and SI~\ref{sec:measurement-of-metrics-from-simulation-data}). We observe a sigmoidal relation between $\theta$ and the dissociation rate $k_-$ for diffusive slip-links, with $\theta$ transitioning from low to high values with increasing $k_-$ (Fig.~\ref{fig:stationary-phenomology}, blue).
	This surprising link between $k_-$ and $\theta$ can ultimately be traced to collective interactions between slip-links: steric hindrance between slip-links drives loops away from the loading site, resulting in ballistic collective motion of slip-links (SI Movie 3b). The lifetime $\tau_{\mathrm{NS}}$ of nested loops  depends on both the velocity of the ballistic movement and on the polymer length $N_m$. We argue below that the increase in $\tau_{\mathrm{NS}}$ due to these two factors quantitatively accounts for the increase of $\theta$ with $k_-$. 	

	We find that the characteristic dissociation rate  at the inflection point of $\theta$ coincides with  the transition from a random polymer to a juxtaposed organization with a cross-diagonal in contact maps ($d_1$ vs. $d_2$ in Fig.~\ref{fig:stationary-phenomology}). In addition, we observe that the equilibrium implementation of diffusive slip-links that bind non-specifically to the DNA (\dParBS) yields $\theta \approx 0$ ($d_3$ in Fig.~\ref{fig:stationary-phenomology}), confirming our previous observation that the loading site is necessary for creating large, nested loops (SI~\ref{sec:diff-slip-links-contact-distribution}). Thus, in the presence of a loading site, diffusive slip-links appear to drive a dynamical transition between phases of weak and strong nesting of DNA-loops, and this transition is crucial to establish the juxtaposed state of the chromosome.
	
	Importantly, in our model with diffusive slip-links, we find that having many nested loops is a necessary, but not sufficient condition to organize the polymer into a juxtaposed state (SI~\ref{sec:Relationship-between-Xc-Xp-and-theta}). These loops also need to propagate into the bulk of the polymer. Indeed, we find that the propagation of diffusive slip-links is a density-driven process: the nested loops only propagate over the full length of the polymer for very high slip-link densities (SI~\ref{sec:universal-scaling-propagation-length}). This can be clearly seen in the binding profiles $p_p(i)$; at low slip-link densities, the binding profile is sharply peaked around the loading site (Fig.~\ref{fig:overview-stationary-results}c, bottom), whereas this peak broadens as we increase the slip-link density (Fig.~\ref{fig:overview-stationary-results}d, bottom).
	
	\subsubsection{Irreversible slip-link binding acts as a kinetic filter for nested loops}\label{subsubsec:mean-field-model}
\begin{figure}
	\includegraphics[width=8cm]{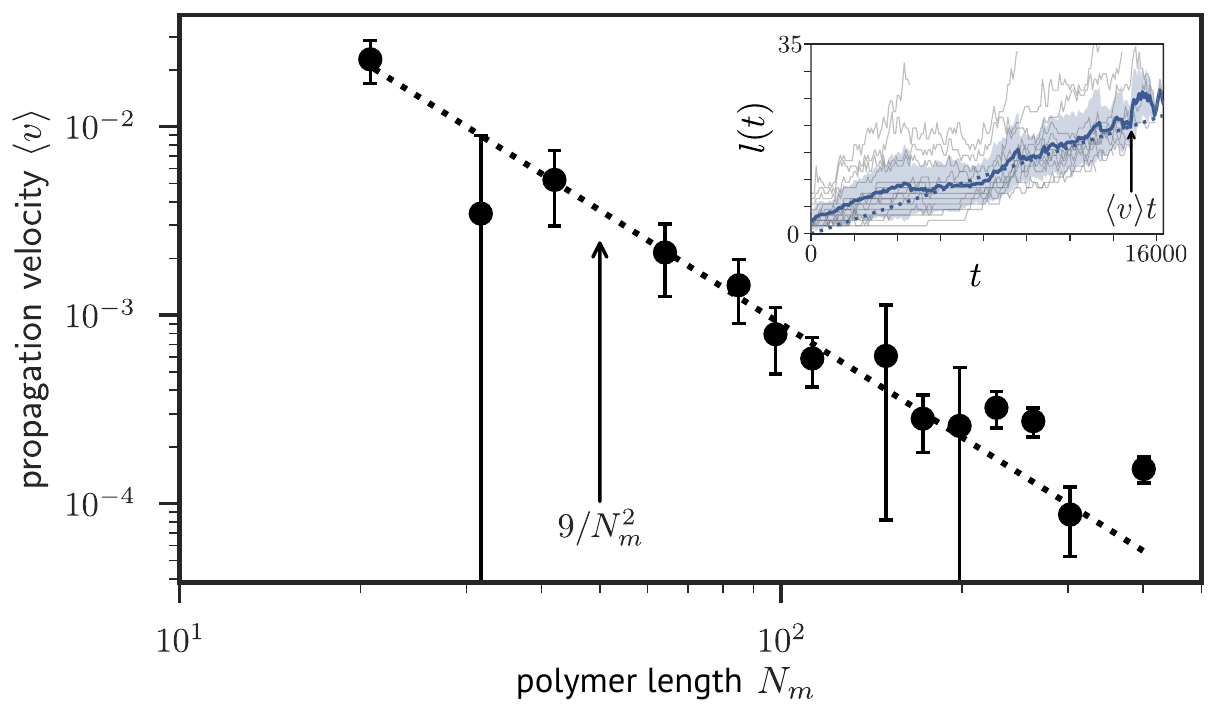}
	\caption{{\small Diffusive slip-links propagate ballistically over a polymer in the juxtaposed state with a velocity $\langle v \rangle \sim 1/N_m^2.$ \textbf{Main panel}: Averaged velocities of tracer slip-links $\langle v \rangle$ (markers) for different system sizes $N_m$ (error bars: twice the standard error in the mean). The dashed line $\langle v \rangle \approx 9k_0/ N_m^2$ is shown together with data computed in the high-density and fast-dissociation regime ($\phi_{p}=0.4, k_- = k_0$).
		\textbf{Inset}: Trajectories of diffusive slip-links $l(t)$ (thin, gray) with ensemble average $\langle l(t) \rangle$ (thick, blue; shaded region is the standard deviation), where $l(t)$ is the distance traversed by a tracer slip-link at  time $t$ after  loading onto the polymer.}
		\label{fig:velocity-of-tracer-slip-links} }
\end{figure}

	To understand the impact of loop lifetime on $\theta$ more quantitatively, we model the loop topology of the  slip-links as $N_p$ independent two-state systems: (i) self-loop (S) with a lifetime $\tau_\mathrm{S}$; (ii) nested loop (NS)  with a lifetime $\tau_{\mathrm{NS}}$. From this mean-field perspective, the fraction of nested loops $\langle \theta \rangle$ is simply the weighted lifetime of a nested loop:
	\[
	\left\langle \theta\right\rangle \approx \dfrac{p_\mathrm{NS}\tau_{\mathrm{NS}}}{p_\mathrm{NS}\tau_{\mathrm{NS}}+p_\mathrm{S}\tau_{\mathrm{S}}},
	\]
	where $p_\mathrm{S}$ and $p_\mathrm{NS}$ are the probabilities for a diffusive slip-link to enclose a self and nested loop respectively after loading to \textit{ori}. The lifetime of a self-loop is $\tau_{\mathrm{S}}\approx k_{-}^{-1}$. We estimate the lifetime of a nested loop by $ \tau_{\mathrm{NS}}\approx \tfrac{1}{2}N_{m} / \left\langle v\right\rangle  + k_-^{-1},$ where $\left\langle v\right\rangle $ is the mean velocity of a slip-link moving through the bulk of the polymer. Since the nested loops propagate ballistically in the high density phase (Fig. \ref{fig:velocity-of-tracer-slip-links}; SI movie 3b), there exists a well-defined velocity $\left\langle v\right\rangle $ that depends on the system size. In particular, we empirically find for $\phi_p = 0.4$ that $\langle v \rangle \approx c k_0 /N_m^2$ where $k_0$ is the slip-link movement attempt rate and $c\approx 9$ (Fig.~\ref{fig:velocity-of-tracer-slip-links}). This $1/N_m^2$ scaling is distinct from the $1/N_m$ scaling observed in the Simple Symmetric Exclusion Process \cite{Derrida2007}, likely due to polymer loop entropy that impedes the movement of slip-links away from the loading site.

\begin{figure}
			\includegraphics[width=8cm]{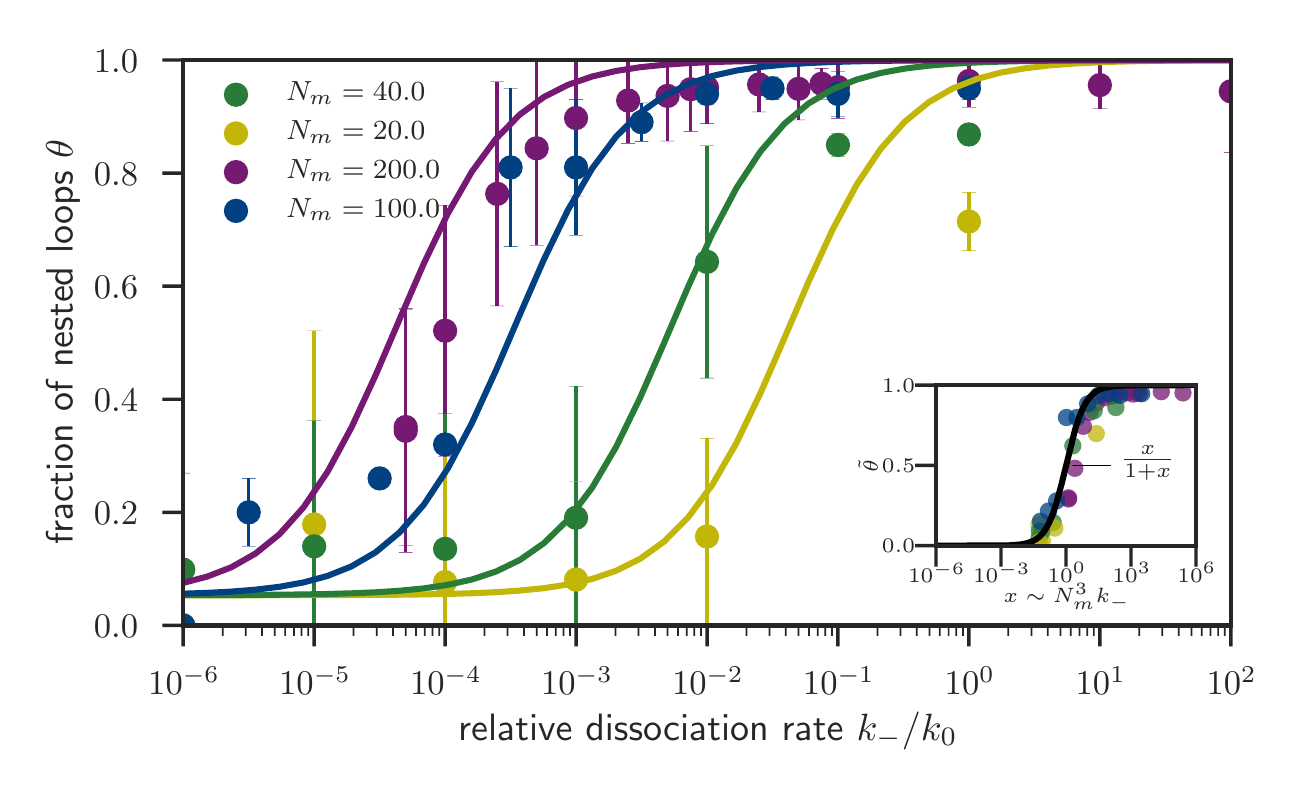}
			\caption{{\small A single master curve approximately describes the fraction of nested loops. \textbf{Main panel}: Loop parameter $\langle \theta \rangle$ (markers) for different system sizes $N_m$ as indicated (error bars: $\sqrt{ \mathrm{var} \theta}$) together with mean-field (Eq.~\eqref{eq:mft-theta}) estimate (solid curves). All data were measured in the high-density regime ($\phi_{p}=0.4$ for $N_m=20,40,200$ and $\phi_{p}=0.38$ for $N_m=100$).
				\textbf{Inset}: The mean-field estimate for $\theta$ (Eq.~\eqref{eq:mft-theta}) collapses the data onto a single master-curve $x/(1+x)$. The reduced parameters are: $\tilde \theta = (\theta - \theta_\mathrm{min})/(\theta_\mathrm{max}-\theta_\mathrm{min}),\theta_\mathrm{max}=1, \theta_\mathrm{min} = \tfrac 1 3 C$ and $x = \tfrac{C}{3-C} ( \tfrac{1}{2}cN_{m}^{3}k_-/k_{0} + 1  )$ with $x \sim N_m^3 k_-$ for $N_m\gg 1$. \label{fig:theta-vs-kmin-data-collapse}} }
		\end{figure}

	In sum, our estimate for $\langle\theta\rangle$ is
	\begin{equation}
	\left\langle \theta\right\rangle  \approx  \dfrac{\tfrac{1}{3}C \left( \tfrac{1}{2}cN_{m}^{3}k_{0}^{-1} + k_-^{-1}  \right)}{\tfrac{1}{3}C \left( \tfrac{1}{2}cN_{m}^{3}k_{0}^{-1} + k_-^{-1}  \right)+(1- \tfrac 1 3 C)k_{-}^{-1}}. \label{eq:mft-theta}
	\end{equation}
		From this, we determine that nested loops start to dominate the loop topology from a characteristic dissociation rate $k_-^\star  \sim N_m^{-3}$.
	The dependency of $\theta$ on $k_-$ reveals that the irreversible loading mechanism functions as a \emph{kinetic filter}: The fast dissociation kinetics filters out self-loops, only allowing nested loops to propagate through the system.
	Since $c$ in Eq.~\eqref{eq:mft-theta} can be determined from a measurement of  $\langle v \rangle$, this form for $\langle \theta \rangle$ does not contain any free fit-parameters, and collapses data for various $N_{m}, k_-$ onto a single master curve (Fig.\,\ref{fig:theta-vs-kmin-data-collapse}).

	\subsection{Motor slip-links are highly effective in organizing the chromosome, even at low densities}
	
	Motivated by recent observations of motor activity of yeast condensin in single-molecule experiments~\cite{Terekawa2017, Ganji2018}, we next explore how such activity impacts the ability of slip-links to organize the chromosome. In our model, motor slip-links are assumed to perform persistent random motion (Fig.~\ref{fig:Microscopic-reactions}, inset). Such persistent slip-links actively extrude loops~\cite{Ganji2018,Sanborn, Fudenberg2016, Brackley2016, Alipour2012}. The active dynamics of motor slip-links is characterized by the  \textit{switching rate} $k_\mathrm{switch}$. For $k_\mathrm{switch} \rightarrow 0$, the motor slip-links  never reverse direction, whereas for $k_\mathrm{switch} \gg k_\mathrm{motor}$ they behave as diffusive slip-links. We find that motor slip-links with small $k_\mathrm{switch}$  organize a system-spanning cross-diagonal (Fig.~\ref{fig:overview-stationary-results}f), whereas the cross-diagonal retracts for increasing  $k_\mathrm{switch}$ (Fig.~\ref{fig:overview-stationary-results}g). 
	
	Overall,  we observe that the persistence of such motor slip-links renders them much more effective at producing the cross-diagonal (Figs.~\ref{fig:overview-stationary-results}f--g). Even for low slip-link densities, a system-spanning cross-diagonal is formed together with an extended binding profile (Fig.~\ref{fig:overview-stationary-results}f). Indeed, motor slip-links can readily drive the chromosome into a state with a high degree of loop nesting. Even in the absence of a loading site ($m_3$ in Fig.~\ref{fig:stationary-phenomology}), motor slip-links efficiently create nested loops.  However, the degree of loop nesting $\theta$ is sensitive to $k_\mathrm{switch}$ (Fig.~\ref{fig:stationary-phenomology}, red). We observe a decline of $\theta$ with increasing $k_\mathrm{switch}$, although $\theta$ remains above 50\% even if the motor slip-link on average switches direction with each step ($k_\mathrm{switch}=k_\mathrm{motor}$). Since a high degree of loop nesting is necessary for establishing the juxtaposed state (SI~\ref{sec:Relationship-between-Xc-Xp-and-theta}), the rate of directional switching must remain sufficiently small  for the motor slip-links to organize the chromosome. 

Importantly, in the absence of a specific loading site, motor slip-links still extrude large loops and efficiently propagate along the polymer, as reflected in the higher contact probability away from the main diagonal (Fig.~\ref{fig:overview-stationary-results}e). However, there is no breaking of translational symmetry by a loading site,  resulting in a leveled time-averaged contact map.
	Interestingly, we observe that systems without a loading site can organize  \textit{transiently} into a juxtaposed state (SI~\ref{sec:motors-without-ori} and SI movies\,7a--b), but the location of the fold diffuses randomly over the polymer.

	\begin{figure}[h!]
		\includegraphics[width=0.73\linewidth]{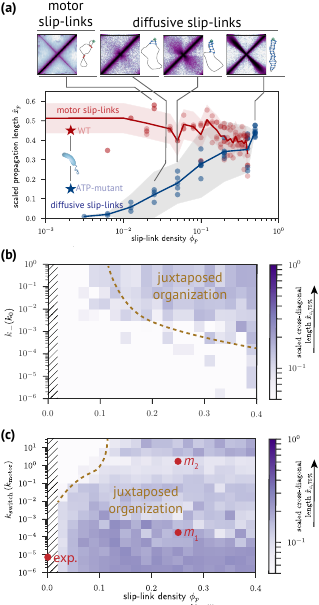}
		\caption{
			{\small Propagation length of slip-links and associated arm--arm juxtaposition \textbf{(a)} Scaled propagation length $\hat x_p \equiv \hat X_p/\tfrac{1}{2}N_m$ as a function of slip-link density $\phi_p$ for diffusive (blue) and motor (red) slip-links. 
			Indicated are also experimental measurements  from~\cite{Minnen2016} for wild-type cells (``WT'', red star) and  mutants whose SMC condensins have suppressed ATPase activity (``ATP-mutant'', blue star) (SI~\ref{sec:empirical-metrics}). 
			\textbf{(b)}: State diagram of scaled cross-diagonal length $\hat x_c  = \hat X_c/\tfrac{1}{2}N_m$ for diffusive slip-links as a function of dissociation rate $k_-$ (in units of $k_0$) and  $\phi_p$. \textbf{(c)}:\,State diagram of scaled cross-diagonal length $\hat x_c$ for motor slip-links as a function of the switching rate $k_{\rm switch}$ (in units of $k_\mathrm{motor}$) and $\phi_p$. Both (b--c) have $N_m=100$. Marker ``exp.'': is an estimate of WT behavior (Table \ref{TABLE}). Markers $m_1$ (SI movie\,10a) and $m_2$ (SI movie\,10b) are respectively at low and high $k_\mathrm{switch}$. Hatched area indicates values  that we did not reach computationally.
We define $\hat x _c$ as the 75$^\mathrm{th}$ percentile of the cross-diagonal contacts (SI~\ref{sec:measurement-of-metrics-from-simulation-data}).}
			\label{fig:stationary-phenomology-2} 
		}
\end{figure}
	
	\subsection{Condensins need motorized movement to efficiently propagate into the bulk of the chromosome}
	
	\textit{In vivo} experiments have demonstrated that condensins propagate far from their loading site~\cite{Minnen2016, Wilhelm2015}. To quantify the distribution of condensins on the chromosome in our model, we measure the extent $\hat X_p$ of slip-link propagation as the standard deviation of the binding profile $p_p(i)$ (Fig.~\ref{fig:overview-stationary-results}h). We eliminate the system-size dependence by considering the scaled propagation length $ \hat x _p = \hat X_p / \tfrac 1 2 N_m$ (SI~\ref{sec:universal-scaling-propagation-length}) as a function of  slip-link density $\phi_p = 2 N_p / N_m$.
	
	The scaled propagation length of diffusive slip-links only approaches the \textit{in vivo} value, $\hat x_p \approx 0.5$ (\cite{Minnen2016, Wilhelm2015} and SI~\ref{sec:empirical-metrics}), when we use slip-link densities~$\gtrsim\!\! 10\%$ in our simulations (Fig.~\ref{fig:stationary-phenomology-2}a, blue). Importantly, this slip-link density would correspond to thousands of condensins on the chromosome, 2--3 orders of magnitude more  than reported  \textit{in vivo}~\cite{Wilhelm2015}. This further illustrates that diffusive slip-links are not efficient at forcing the nested loops into the bulk of the polymer at low densities. In contrast, motor slip-links propagate over the full length of the polymer at all slip-link densities we considered (red data in Fig.~\ref{fig:stationary-phenomology-2}), as observed experimentally in cells (\cite{Minnen2016, Wilhelm2015} and SI~\ref{sec:empirical-metrics}).
	
	Our results are summarized in a ``state diagram'' (Figs.~\ref{fig:stationary-phenomology-2}b--c), indicating the scaled extent $\hat x_c = \hat X_c/\tfrac 1 2 N_m$ of the cross-diagonal in contact maps. 
	\textit{Diffusive} slip-links require both fast dissociation kinetics as well as a high slip-link density to bring the polymer into the juxtaposed state (Fig.~\ref{fig:stationary-phenomology-2}b). For  high densities of \textit{motor} slip-links $\phi_p \gtrsim 10\%$ and low $k_\mathrm{switch}$, the  slip-links readily juxtapose the DNA polymer (e.g. $m_1$ in Fig.~\ref{fig:stationary-phenomology-2}c). Contrarily, for increasing $k_\mathrm{switch}$, motor slip-links  antagonize each other's translocation (e.g. $m_2$ in Fig.~\ref{fig:stationary-phenomology-2}c), impeding the collective propagation of  slip-links away from the loading site, thereby resulting in a reduced $\hat x_c$ (compare $m_1$, $m_2$ in Fig.~\ref{fig:stationary-phenomology-2}c; SI movies\,10a--b). In the limit $k_\mathrm{switch} \gg k_\mathrm{motor}$,  motor slip-links effectively behave as diffusive slip-links with enhanced unbinding kinetics, placing them in the fast dissociation regime (data for $k_\mathrm{switch} \gtrsim 10 k_\mathrm{motor}$ in Fig.~\ref{fig:stationary-phenomology-2}c).
In contrast, for $\phi_p \lesssim 10\%$,  motor slip-links only require  $k_\mathrm{switch}$ to be sufficiently low. We estimate that wild-type condensin is indeed in this slow switching regime (``exp.'' in Fig.~\ref{fig:stationary-phenomology-2}c).	
	In sum, our simulations indicate that condensins at physiological densities can drive nested loops into the bulk of the polymer, crucial for establishing arm--arm alignment,  only if they perform motorized, persistent motion.

	\subsection{SMC condensin requires motor activity to rapidly re-organize the chromosome}
		
\begin{figure}
			\includegraphics[width=\linewidth]{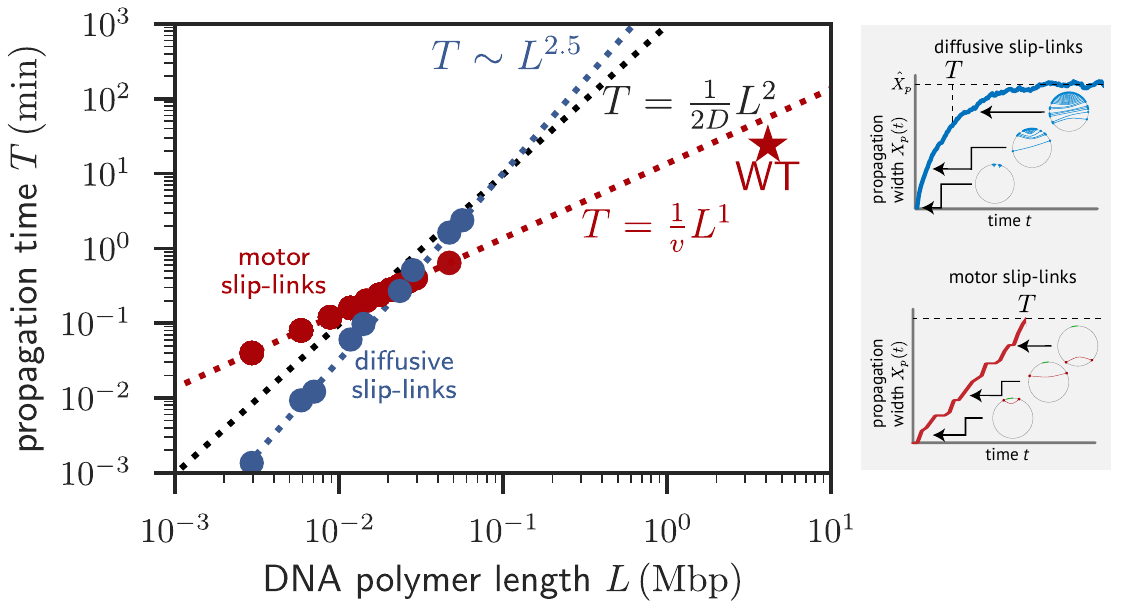}
			\caption{{\small Propagation dynamics of slip-links and comparison with wild-type cells. For diffusive slip-links (blue), we quantify the propagation time-scale $T$ by an exponential fit to the propagation length $X_p (t)$ (legend, top). For motor slip-links (red), we  computed the average translocation velocity $v$ and define $T = L/v$ (legend, bottom). %
						Diffusive slip-links exhibit sub-diffusive scaling  $T \sim L^{2.5}$ (blue, dashed); diffusive scaling $T\sim L^2$ is shown for comparison (black, dashed).  The extrapolation of the simulated data for motor slip-links (red) is in accord with  \textit{in vivo} data (``WT'', see SI~\ref{sec:empirical-metrics}). Simulation units were converted to real units as described in SI~\ref{sec:empirical-metrics}.}
				\label{fig:dynamic-phenomenology}
			}
	\end{figure}

	SMC induction experiments revealed that condensin can propagate from the loading site into the bulk of the DNA, thereby organizing an entire bacterial chromosome in a timespan of only $ T_\mathrm{WT} \approx 24 \unit{min}$ (SI~\ref{sec:empirical-metrics} and~\cite{Wang2017a}). Based on these experiments, we estimate that condensins translocate away from their loading site with a velocity of $\approx 300 \unit{nm/s}$. To understand the origin of these remarkably fast dynamics, we compute time-traces $X_p (t)$ of the width of the slip-link binding profile for our minimal models. From these traces, we extract  a typical propagation time $T$ for slip-links to establish a steady-state binding profile on a DNA polymer of physical length $L = a N_m$ (Fig.~\ref{fig:dynamic-phenomenology}, inset), where  $a\approx 50 \unit{nm}$ is the size of one monomer (Table~\ref{TABLE}).

	The propagation time of diffusive slip-links scales strongly with DNA length: $T \sim L^x$, with $x\approx 2.5$ (Fig.~\ref{fig:dynamic-phenomenology}, blue). This scaling differs from simple diffusive motion ($x=2$, black in Fig.~\ref{fig:dynamic-phenomenology}), which we attribute  to the loop-entropic forces that impede slip-link movement away from their loading site. Based on this observed scaling of $T$, we estimate that diffusive slip-links propagate several orders of magnitude slower over the DNA than observed in live cells (Fig.~\ref{fig:dynamic-phenomenology}, ``WT'').  In  contrast, the propagation time of motor slip-links exhibits ballistic scaling, $T = L/v$ (Fig.~\ref{fig:dynamic-phenomenology}, red), where $v$ is the effective translocation velocity.

	Interestingly, our model prediction of motor slip-links, $T = v L $, is remarkably close to the observed propagation time \textit{in vivo} (Fig.~\ref{fig:dynamic-phenomenology}, ``WT''). Indeed, recent single-molecule experiments have revealed that yeast condensin can extrude DNA loops with a velocity of up to $425 \unit{nm/s}$~\cite{Ganji2018}. These data combined with our simulations, strongly indicate that rapid re-organization of the chromosome by SMC condensin requires fast and active loop extrusion.
	
	\subsubsection{DNA relaxation dynamics can limit slip-link velocity}
	
\begin{figure}
		\centering
		\includegraphics[width=8cm]{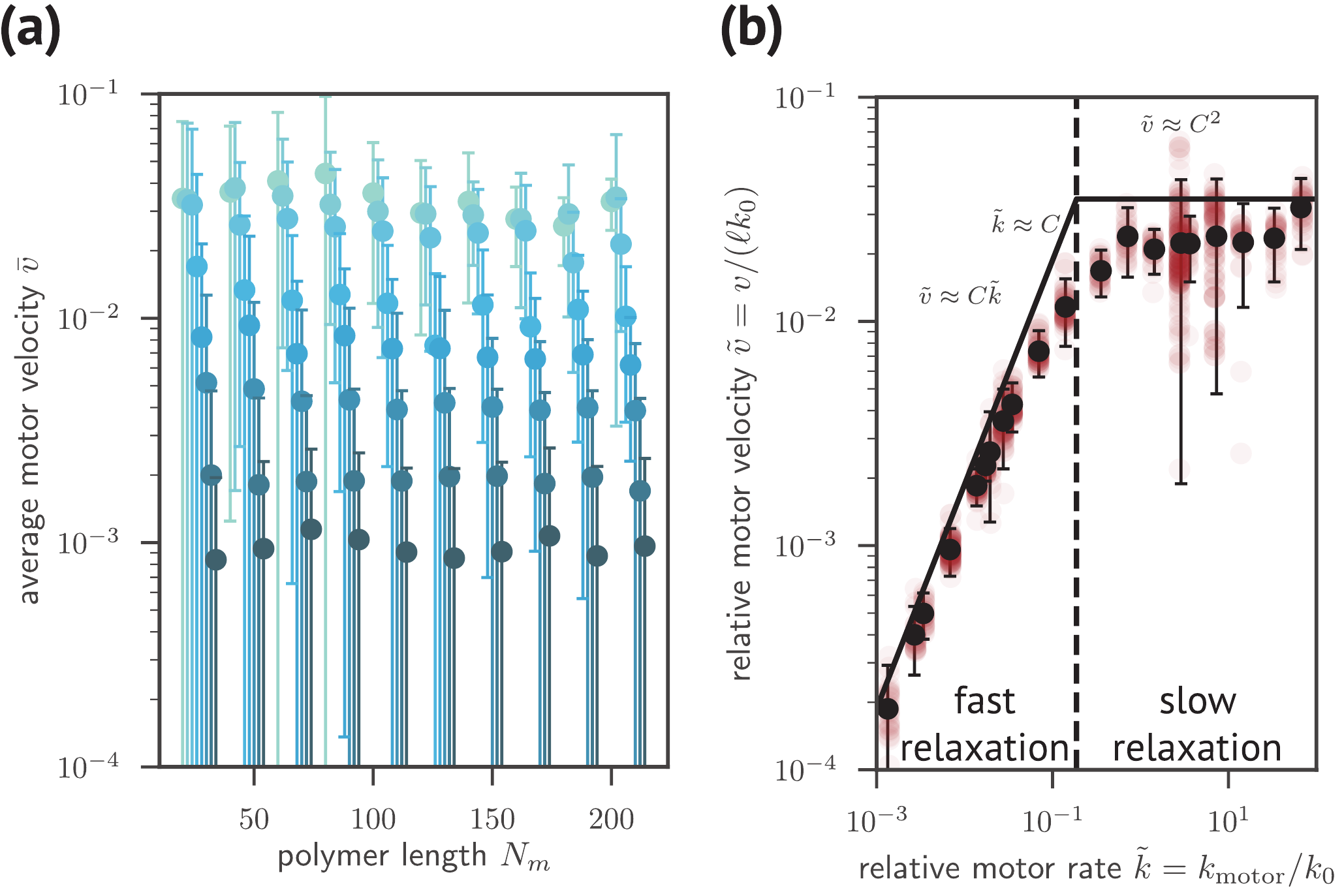}
		\caption{{\small DNA relaxation dynamics limits the maximum velocity of motor slip-links. \textbf{(a):} The average motor velocity (error bars: standard deviation) for various translocation attempt rates (turquoise to dark blue, $\kMotor = 100\ldots 0.01$) and various system sizes $N_m$. For clarity, we shifted the datapoints for decreasing $\kMotor$; the value of $N_m$ used is always the datapoint corresponding to $\kMotor=100$ (turquoise). \textbf{(b):} Rescaled motor velocity as a function of relative motor rate for the data shown in panel a). Propagation of wild-type condensins is within the ``fast relaxation'' regime ($\kMotorRed=6.8\times 10^{-6}$) using data from Table \ref{TABLE}.}}
		\label{fig:motor-dynamics-combined}
	\end{figure}

	What is the relationship between the slip-link translocation attempt rate $\kMotor$, the monomer diffusion attempt rate $k_0$, the polymer size $N_m$, and the effective motorized slip-link translocation velocity $v$? To answer this question, we distinguish two regimes: a \textit{fast} relaxation regime $\kMotor \ll k_0$ and a \textit{slow} relaxation regime $\kMotor \gg k_0$. In both  regimes, for a stiff slip-link to make a step, the two polymer bonds in the direction of movement need to be parallel (Fig. S\ref{fig:illustration-counting-argument}a). If we denote the prior probability of observing these two polymer bonds to be parallel by $C \approx 3/16$ (Fig. S\ref{fig:illustration-counting-argument}b), then the effective rate of bond--bond alignment is $k_0^\mathrm{eff} \approx C k_0$. In our estimate for the rate $k_0^\mathrm{eff}$ we neglect the force that motor slip-links might exert on these bonds, because \emph{in vitro} experiments indicate that the stalling force of yeast condensin is very small \cite{Ganji2018}. Hence, the characteristic rate $\kMotor^\star$ that sets the transition from the fast relaxation to the slow relaxation regime occurs at $\kMotor^\star = k_0^\mathrm{eff}$. Consistent with the prediction that $\kMotor$ only depends on \textit{local} kinetics, our data shows that $\kMotor$ is independent of the system size (Fig. \ref{fig:motor-dynamics-combined}a).
	
In the fast relaxation regime, the rate-limiting factor is $\kMotor$.  In this case, the velocity $v$ of slip-links is approximated by $v_\mathrm{fast}(\kMotor) \approx C \ell \kMotor$. In the slow relaxation regime, $\kMotor > \kMotor^\star$, polymer relaxation becomes the rate-limiting factor, so that $v_\mathrm{max} \approx v_\mathrm{fast}(\kMotor^\star)  \approx C^2 \ell k_0 $.
This argument suggests that the scaling form of dimensionless variables $\tilde v \equiv v / (\ell k_0)$,  $\kMotorRed \equiv \kMotor / k_0 $ will collapse the data of $\kMotor, v, k_0$ onto a universal curve $\tilde v = C \kMotorRed$ for $\kMotorRed < C$ and $\tilde v = C^2$ for $ \kMotorRed \geq C$. Indeed, our numerical data is well-described by this scaling form (Fig.\,S\ref{fig:motor-dynamics-combined}b). By combining \textit{in vitro} with \textit{in vivo} empirical data, we estimate that SMC condensin in \textit{B. subtilis} is well within the fast relaxation regime $\kMotor < 10^{-5} k_0 \ll k_0 $.

	\section{Discussion}
	Our computational framework reveals the basic physical requirements for condensins to collectively organize the bacterial chromosome as observed in live cells~\cite{Wang2015a, Wang2017a, Marbouty2015, Le2013}. In the presence of a specific loading site and with physiologically relevant numbers of condensins,  we find that motor activity is required to robustly and rapidly generate a system-size spanning juxtaposition of the chromosomal arms. In contrast, purely diffusive condensins would require more kinetic fine-tuning  and unphysiologically high copy numbers to organize the chromosome.
	
	Our minimal model for the action of SMC condensin as a motor slip-link accounts for several key observations, including the rapid development of the juxtaposed state~\cite{Wang2017a} and the crucial role of the \ParBS nucleoprotein complex as a specific loading site~\cite{Wang2015a, Wang2017a, Gruber2009, Marbouty2015}. Without a well-defined loading site (\dParBS), motor slip-links still transiently organize the polymer into a juxtaposed state. However, the chromosomal fold diffuses along the chromosome, resulting in a structureless time-averaged contact map with enhanced long-range contacts (Fig.~\ref{fig:overview-stationary-results}e and SI~\ref{sec:motors-without-ori}). These theoretically predicted behaviors may account for observations in \textit{E. coli}, where the action of the SMC complex MukBEF does not appear to involve an exclusive loading site~\cite{Lioy2018}. Hi-C maps of \textit{E. coli} do not display a cross-diagonal, but rather an elevated contact probability at large length-scales~\cite{Lioy2018}, in line with our simulations (Fig.~\ref{fig:overview-stationary-results}e and SI~\ref{sec:motors-without-ori}). Indeed, our model predicts that motor slips-links efficiently create nested loops ($m_3$ in Fig.~\ref{fig:stationary-phenomology}) even in the absence of a loading site, resulting in enhanced long-range contacts (Fig.~\ref{fig:overview-stationary-results}e and Fig.~S\ref{fig:contact-distribution-motor-wo-loading-site}). Although the function of such a change in polymer organization is unclear, local chromosome organization by SMC proteins  has been linked to transcription regulation in various bacteria such as \textit{E. coli} and \textit{C. crescentus}~\cite{Lioy2018, Le2013}.
	
	Experiments of SMC condensin propagation in \textit{B. subtilis} suggest that two condensin complexes might link together in a hand-cuff topology, with each of the two condensins in the dimer actively extruding a separate DNA duplex~\cite{Wang2017a, Tran2017}. In our model, motor slip-links extrude DNA in a symmetrical fashion, as expected for condensins in a hand-cuff configuration: both sides of a slip-link move over a separate DNA duplex with the same translocation rate.  However, in recent \textit{in vitro} assays, single yeast condensin complexes actively extrude DNA loops asymmetrically: one end of the complex appears anchored at a DNA locus, while the opposite end actively translocates over DNA~\cite{Ganji2018}. Interestingly, contact maps of such asymmetric motor slip-links contain a star-shaped pattern around the loading site (SI~\ref{sec:asymmetric-loop-extrusion}), a feature that is also visible in Hi-C maps of \textit{B. subtilis}~\cite{Wang2015a, Marbouty2015}. This suggests that there is at least some fraction of condensins performing asymmetric translocation. An equally tantalizing explanation is that the movement of one of the condensins in a dimer is impeded by other DNA-bound factors, thereby forcing a condensin-dimer to propagate asymmetrically. Indeed, there is growing evidence that the movement of SMC complexes can be antagonized by oncoming transcription factors~\cite{Wang2015a, Tran2017, Wang2017a}.
	
	It has also been suggested that the cylindrical geometry of many bacteria, combined with DNA--cell-pole tethering, facilitates chromosome organization~\cite{Buenemann2010, MathiasBuenemann2011}. However, our simulations indicate that the effect of confinement alone on chromosome organization is  weak, and hence cannot be solely responsible for the juxtaposed organization observed in live cells (SI~\ref{section:cross-diagonal-from-cell-asymmetry}). This is in line with the observation that mutants lacking SMC condensin, but with the \textit{ori}-proximal loci still tethered to a cell-pole, also lack the cross-diagonal~\cite{Wang2017a}.
	
	Our simulations further indicate that even purely diffusive slip-links with fast dissociation kinetics can induce arm-arm alignment, but at low slip-link densities this organized state remains localized near the  loading site (SI~\ref{section:relaxatin-juxtaposed-organization}).
	For a given number of condensins (3--30 per chromosome \cite{Wilhelm2015}), however, the slip-link density depends on DNA length. Therefore, we expect that a physiological number of non-motorized condensins can organize a mini-chromosome or plasmid of tens of microns in length. Indeed, plasmids are known to bind SMCs~\cite{Kumar2006} and contain \ParBS nucleoprotein complexes that could act as a condensin loading site~\cite{Funnell2016}.

	Finally, our computational model can be used to unravel the function of juxtaposed organization in faithful chromosome segregation~\cite{Wang2013a,Wang2014, Hirano2016, Gruber2009, Gruber2014, Marbouty2015, Nolivos2013a}. More broadly, we provide a framework to elucidate the role of  loop-extruding enzymes  and ATPases~\cite{Brackley2018, Brackley2016, Hirano2016, Gruber2014a, Lioy2018} on chromosome organization.

\section*{Author Contributions}
C.A.M. and C.P.B. designed research; C.A.M. performed research; C.A.M. and C.P.B. analyzed data; C.A.M. and C.P.B. wrote the paper.

\section*{Data Accessibility}
Original data of contact maps, polymer configurations and slip-link dynamics displayed in this paper are available upon request in the form of HDF5 data files.

\section*{Funding Statement}
This work was supported by  	the German Excellence Initiative via the program ``NanoSystems Initiative   Munich''   (NIM) and   the Deutsche Forschungsgemeinschaft   (DFG)  Grants TRR174 and GRK2062/1.

	\section*{Acknowledgments}
	We kindly thank S. Gruber, D. Rudner, E. Frey, J. Messelink, T. Krueger, F. Mura, F. Gnesotto, L. van Buren, G. Dadushnavili, and A. Le Gall for their insights on this work and stimulating discussions.

	\bibliography{references}

	\newpage
	
	\setcounter{section}{0}
	\setcounter{figure}{0}

\onecolumngrid
\newpage

\begin{appendices}
	\textbf{Supplementary Information to "Bacterial chromosome organization by collective dynamics of SMC condensins", C.A. Miermans \& C.P. Broedersz, 2018}

	 \captionsetup[figure]{labelfont={bf},name={Fig.~S\!\!},labelsep=period}
	\section{Kinetic Monte-Carlo algorithm\label{sec:Explanation-of-KMC}}
	
\begin{figure}[b!]
		\begin{centering}
			\includegraphics[height=8cm]{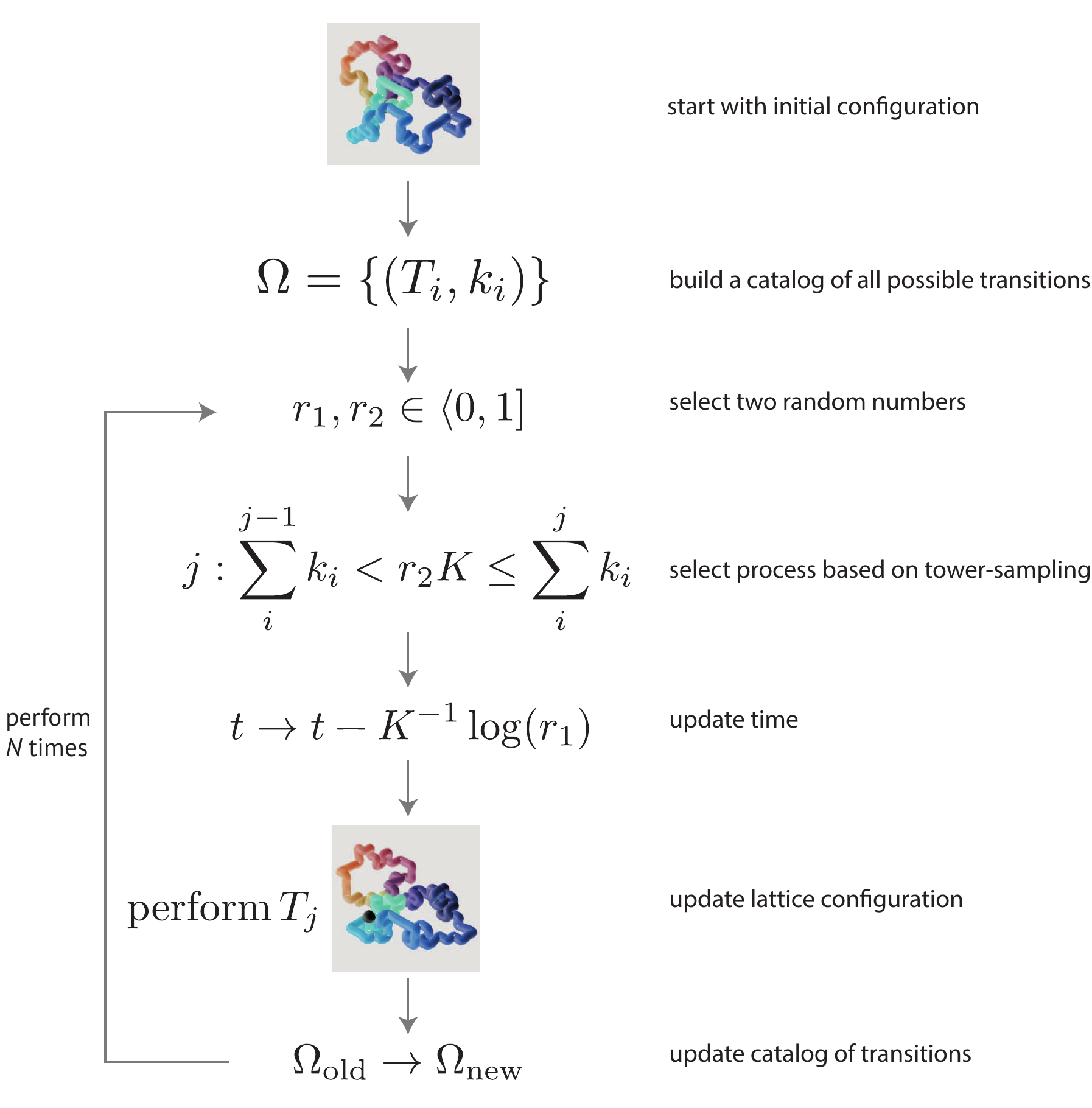}
			\caption{\textbf{Schematic illustration of KMC algorithm} More details are found in section~\ref{sec:Explanation-of-KMC}. \label{fig:schematic-kmc-algorithm} }
		\end{centering}
	\end{figure}

	Our lattice Kinetic Monte-Carlo (KMC) framework employs a Gillespie-type algorithm \cite{Voter}. This KMC algorithm is schematically illustrated in Fig.\,S\ref{fig:schematic-kmc-algorithm}, and consists of the following steps:
	\begin{enumerate}
		\item Construct initial configuration of the DNA polymer $\{ \v r_i \}$
		\item Build a rate catalog $\Omega = \{(T_i, k_i)\}$ of all possible transitions $T_i$ and their associated rates $k_i$, with all transitions $T_i$ assumed to be Poissonian. Next, enter  the loop consisting of steps (a)--(e):
		\begin{enumerate}
			\item Randomly select one of the transitions $T_j$ in $\Omega$. The probability to perform transition $T_i$ is $k_i /\sum_i k_i$, which we implemented using tower sampling \cite{Krauth2006}.
			\item Update the KMC time $t \rightarrow t+\Delta t, \Delta t = - \log r / K$, where $r$ is a uniformly sampled random number $r \in \langle 0,1 ]$ and $K = \sum_i k_i$ is the total rate of the system \cite{Voter}.
			\item Perform  transition $T_j$ that was selected in (a), which can affect the DNA polymer or the particles interacting with the polymer.
			\item Update the entire rate catalog of possible transitions $\Omega$ based on the transition $T_j$ that was just performed. For more details, see figure~\ref{fig:schematic-kmc-algorithm}. 
			\item Return to step (a).
		\end{enumerate}
		
	\end{enumerate}

\begin{figure}[t!]
		\begin{centering}
			\includegraphics[width=7cm]{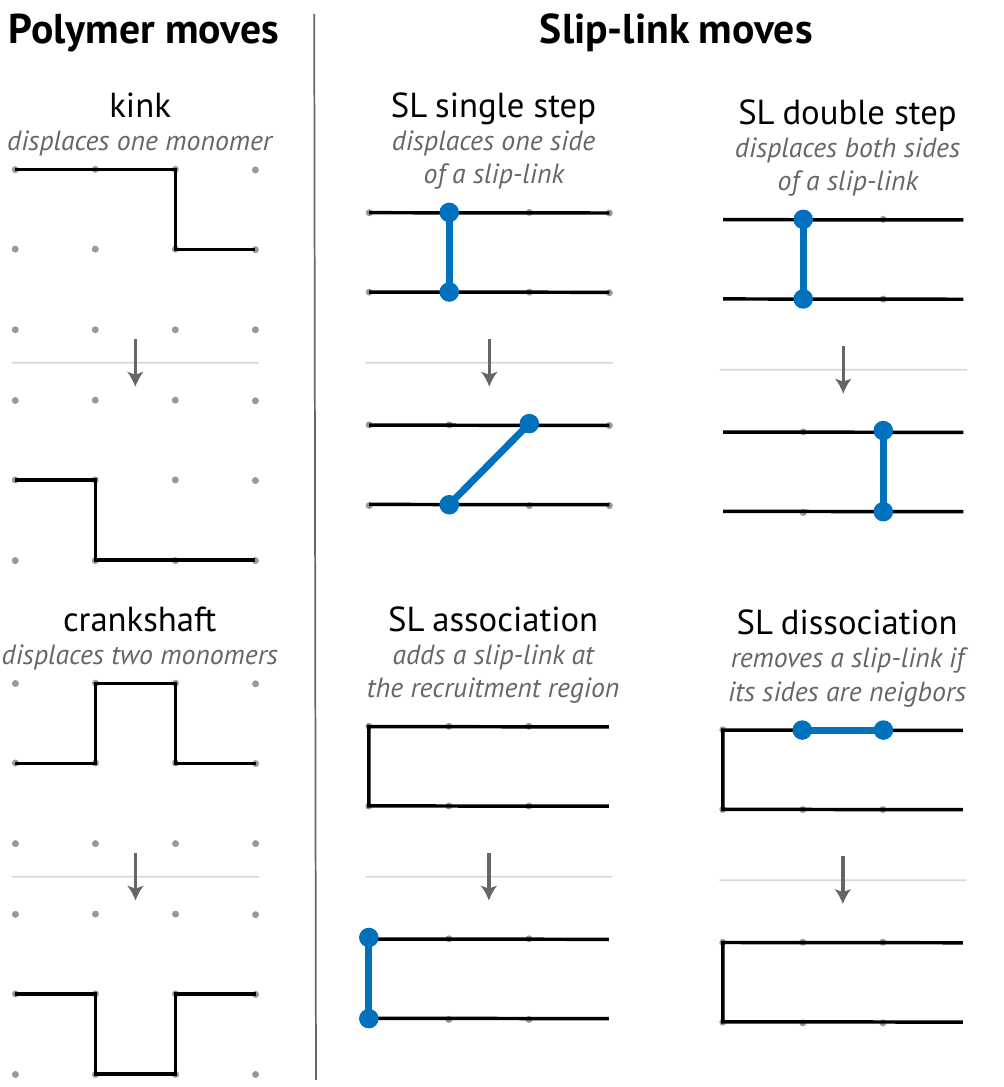}
			
			\caption{ \textbf{Schematic illustration of lattice KMC moves} Left column: Polymer dynamics is described by the Verdier--Stockmayer move-set \cite{Doi1986}. Right columns: Slip-link moves include association to the DNA polymer and movement along the polymer. \label{fig:SI-kmc-moves} }
		\end{centering}
	\end{figure}

	Every iteration of the KMC loop (see Fig. S\ref{fig:schematic-kmc-algorithm}) requires an update of the rate catalog $\Omega$. For computational efficiency we only update the part of the rate catalog $\delta \Omega_j \in \Omega$ that is possibly affected by the previous transition $T_j$. Let $\Theta_j$ be the set of coordinates of monomers that are displaced by the transition $T_j$. We update all  coordinates $\v q$ in the rate catalog  that satisfy $|\v p - \v q| \leq m, \v p \in \Theta_j$, since our move set displaces particles by at most $m = 2$ lattice points (Fig.\,S\ref{fig:SI-kmc-moves}, ``crankshaft'').

	The lattice KMC moves that we employ are illustrated in Fig.\,S\ref{fig:SI-kmc-moves}. Importantly, this move-set with local moves for the polymer dynamics is guaranteed to give rise to  Rouse dynamics \cite{Doi1986}.
	
	Our simulation parameters and units are clarified in Table \ref{TABLE}. Because cytosolic diffusion of condensin is fast compared to diffusion
	of condensin over dsDNA \citep{Stigler2016a}, we assume a regime
	in which the condensin binding rate is much higher than the slip-link movement attempt
	rate, $k_{+}\gg k_{0}$.  We implemented loading to the \ParBS region \citep{Minnen2016} by binding slip-links only to a
	predetermined loading site (\textit{ori}) on the polymer. Slip-links can, however, unbind
	anywhere on the DNA, as long as the slip-link encircles a loop of only one lattice
	width (Fig. S\ref{fig:SI-kmc-moves}). 
		\begin{table}
		\caption{The parameters used in the simulations and comparisons with WT data.\label{TABLE}}
		\newcolumntype{R}{>{\raggedleft\arraybackslash}X}%
		\begin{tabularx}{\textwidth}{lR | lR}
	\hline\hline			\vspace{1pt}
			\vspace{-10pt}	& & & \\
\textbf{Quantity} & \textbf{Expression} & \textbf{Value} & \textbf{Reference}  \\ 
			\hline
			\small Lattice constant (persistence length of dsDNA) & $\ell_0$ & 50 nm  & \citep{Smith1992} \\
			Size of slip-link & $25-50 \unit{nm}$ &  set to $\ell_0$  & \cite{Eeftens2016, Burmann2017} \\
			Size of \textit{B. subtilis} genome & $L_\mathrm{gen.}$  &  $4 \unit{Mbp}$  & \citep{Minnen2016} \\
			Length of \textit{B. subtilis} genome in sim. units & $N_{m,\mathrm{WT}} = L_\mathrm{gen.}/\ell_0$ &  $28\cdot 10^3$  & rest of table \\
			Radius of dsDNA & $b$ & $2\unit{nm}$  & \cite{Moran2010a} \\
			Thermal energy & $k_B T$ & $ 4.2 \unit{pN \, nm}  $& \cite{Moran2010a} \\
			Cytosolic viscosity   &  $\eta$ & $1 \unit{mPa/s}$ & \cite{Bicknese1993, Swaminathan1997, Ando2010} \\
			Diff. coeff. of dsDNA-bound cohesin & $D$ & $1 \unit{\mu m^2/s}$ & \cite{Stigler2016a} \\
			Diffusion time of slip-link over monomer     &$\tau_\mathrm{diff. slip-link} \approx {\ell_0^2}/{2 D}$  & $1250 \unit{\mu s}$ & rest of table \\
			Loop extrusion velocity of yeast condensin       & $v$  & $425 \unit{nm/s}$ & \cite{Ganji2018}  \\
			Translocation time of condensin over a monomer      & $	\tau_\mathrm{motor. slip-link} \approx {\ell_0}/{v}$  & $0.12 \unit{s}$ & rest of table \\
			Yeast condensin switching rate        & $k_\mathrm{switch}$ & $222 \unit{min^{-1}}$ & SI\,\ref{sec:empirical-metrics}  \\
			Relative SMC propagation length, WT & $\hat x_\mathrm{SMC, WT}$ & $45\%$ & SI \ref{sec:empirical-metrics} \\ 
			Relative SMC propagation length, -ATP & $\hat x_\mathrm{SMC, ATP_-}$ & $13\%$ & SI \ref{sec:empirical-metrics} \\ 
			Monomer relaxation time & $	\tau_\mathrm{mon.} = \small\frac{2}{\pi} {\ell_0^2 b \eta}/{k_B T}$ &  $0.8 \unit{\mu s}$  & \cite{Rubinstein2003,Tothova}  \\
			Number of condensins per chromosome & $N_{p, \mathrm{WT}}$ & 3--30 & \cite{Wilhelm2015} \\
			SMC density per simulation monomer & $\phi_{p,\mathrm{WT}} = \frac{2N_{p, \mathrm{WT}}}{N_{m,\mathrm{WT}}}$ & $10^{-4}-10^{-3}$ & rest of table \\

			\vspace{-7pt}  & & & \\
			\hline\hline 					
		\end{tabularx}
	\end{table}

	\section{Measurement of metrics  from empirical data}\label{sec:empirical-metrics}
	
	\subsection*{Motor direction switching time in yeast condensin}
	
	\textit{In vitro }DNA curtain experiments have shown that a fraction
	of condensins can reverse their active motion within a time $\tau_r$
	\citep{Terekawa2017}. In these experiments, the condensins
	can only be monitored over a maximum distance $L_{\mathrm{assay}}=16.49\unit{\mu m}$. The time a condensin can be observed to move with a constant velocity $v\approx20\unit{nm/s}$
	is, therefore $t^{\star}=L_{\mathrm{assay}}/v\approx825\unit{s}$
	\citep{Terekawa2017}. In \citep{Terekawa2017}, the probability for
	the motor to be reversed at least once within this time
	$t^{\star}$ has been measured to be $p_r(t^{\star})\approx6\%$
	. Assuming Poissonian statistics, the probability of the condensin
	having reversed within a time $t^\star$ is
	\[
	p_r(t^{\star}) = 1 - \exp(-t^{\star}/\tau_r),
	\]
	from which we estimate a typical
	switching time for yeast condensin $\tau_r\approx222\unit{min}$. Based on a typical velocity of $v\approx 20\unit{nm/s}$~\cite{Ganji2018}, we estimate that a yeast condensin travels on average a distance $L_{\mathrm{max}}=v\tau_r\approx 266\unit{\mu m}\approx785\unit{kpb}$ before switching direction. Crowding on the DNA by other proteins \textit{in vivo} could affect these estimates.
	
	\subsection*{Measuring propagation length of condensin in \textit{B. subtilis}}
\begin{figure}[b!]
		\begin{centering}
			
			\includegraphics[width=8cm]{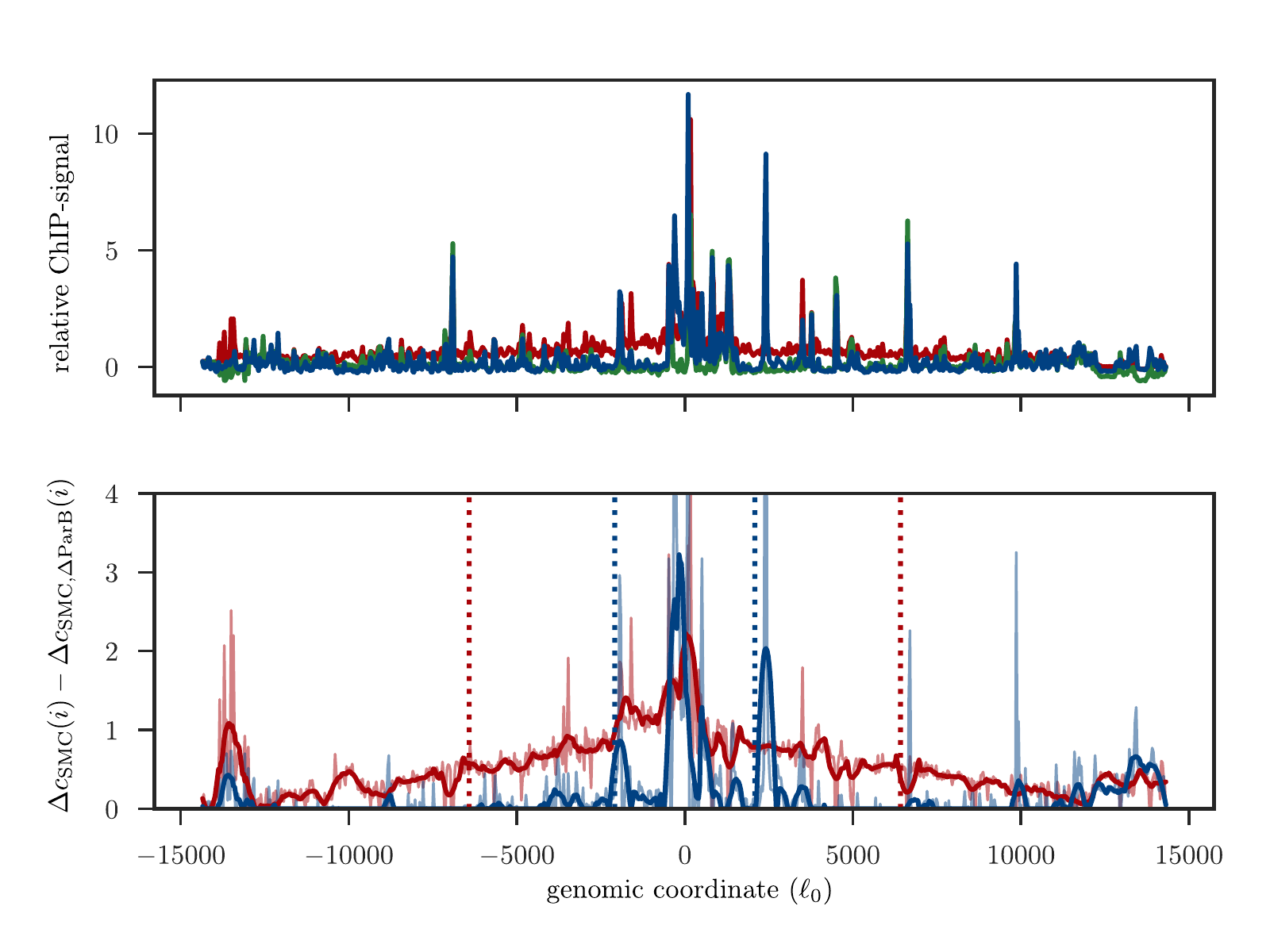}
			
		\end{centering}
		\caption{\textbf{Determination of typical SMC propagation width from SMC ChIP-seq data.}. \textbf{Top}: Raw relative ChIP-seq data $ c_\mathrm{SMC}(i) - c_\mathrm{SMC}(\tfrac{1}{2}N_m)$ (i.e. we subtracted the ChIP-signal at \textit{ter}) of three strains from
			\citep{Minnen2016}: ``WT'' (wild-type, red), ``$\mathrm{ATP}_-$'' (mutant with
			strongly suppressed ATP hydrolysis, blue) and $\Delta$ParB (mutant
			lacking ParB, green).  \textbf{Bottom}: We use the $\Delta$ParB signal as a baseline for the SMC ChIP-signal. The WT and $\mathrm{ATP}_-$ signals with the $\Delta$ParB
			signal subtracted (thin curves) and with an additional smoothing using
			a Savitzky-Golay filter (window: 21, order: 2) (thick curves). The typical width of these graphs was then defined
			as $\sqrt{\protect\var [ c_\mathrm{SMC}(i) ]}$ (dotted vertical lines). Lengths
			are in units of polymer bond lengths $\ell_{0}=50\unit{nm}$ (see Table
			\ref{TABLE}). ChIP-seq data were taken from the ArrayExpress database at EMBL-EBI (www.ebi.ac.uk/arrayexpress) under accession number E-GEOD-76949. \label{fig:chip-seq-data-from-minnen}}
	\end{figure}

	We analyzed ChIP-seq data from \citep{Minnen2016} of the strains ``WT''
	(wild-type), ``$\mathrm{ATP}_-$''  (mutant with strongly suppressed ATP hydrolysis)
	and $\Delta$ParB (mutant without ParB). As a baseline for the SMC
	signal, we used the $\Delta$ParB data for both the WT and $\mathrm{ATP}_-$ strains
	and computed the difference in their SMC ChIP-Seq signals $c_\mathrm{SMC}(i)$ (Fig.\,S\ref{fig:chip-seq-data-from-minnen}),
	\begin{eqnarray*}
		\Delta c_{\mathrm{SMC,WT}}(i) 	= & c_{\mathrm{SMC,WT}}(i)-c_{\mathrm{SMC,\Delta ParB}}(i) \\
		\Delta c_{\mathrm{SMC,ATP_-}}(i)			= &c_{\mathrm{SMC,ATP_-}}(i)-c_{\mathrm{SMC,\Delta ParB}}(i) .
	\end{eqnarray*}
	We then smoothed the data with a Savitzky-Golay-filter of window 21 and order 2 and extracted typical widths $\hat X_\mathrm{SMC}$, defined as the standard deviation of the smoothed ChIP profiles. We computed the typical widths $\hat X_\mathrm{SMC, WT} \approx7000\ell_{0},\hat X_\mathrm{SMC, ATP_-} \approx2000\ell_{0}$---or, when scaled by the system size $\hat x_\mathrm{SMC, WT} \approx 13 \%,\hat x_\mathrm{SMC, ATP_-} \approx 45\%$.
	
	\subsection*{Measuring condensin propagation timescale}
	
\begin{figure}[t!]
		\includegraphics[width=6cm]{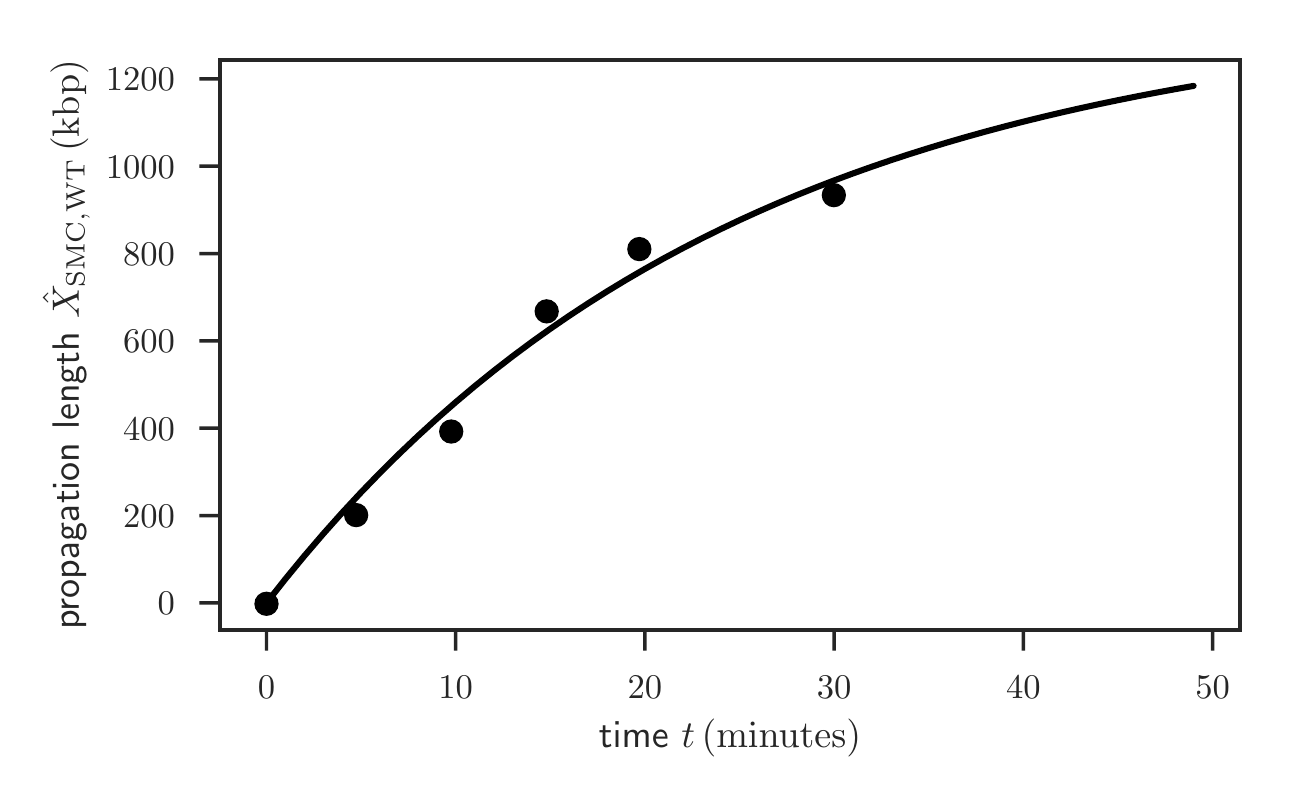}
		
		\caption{The ChIP-seq data of the SMC condensin propagation width has a typical timescale of $\approx 24 \unit{min}$. Fit parameters $\tau \approx 24\unit{min}, \mathrm{amplitude} \approx 1368 \unit{kbp}$  were found by fitting the data from \cite{Wang2017a} with an exponential.  \label{fig:typical-propagation-timescale-of-smc}
		}
	\end{figure}

	Recent \textit{in vivo} experiments have been performed in which ChIP-seq data was measured at various time-points after induction of SMC condensin \cite{Wang2017a}. From these ChIP-seq data, the typical width $\hat X _ p (t)$ of condensin propagation  was measured as a function of time. Since the curve of $\hat X _ p (t)$ was well approximated by an exponential curve, we extracted a typical timescale of $\approx 24 \unit{min}$ from these curves (Fig.\,S\ref{fig:typical-propagation-timescale-of-smc}).
	
	\section{Measurement of metrics from simulation data}\label{sec:measurement-of-metrics-from-simulation-data}
	
	\subsection*{Measuring the loop network topology}
	
	We quantify the loop network topology using an order parameter $\theta \in [0,1] $ that measures the fraction of nested loops. This order parameter is defined for a slip-link--polymer network, so that we can measure $\theta$ for each time-point $t$, i.e. $\theta(t)$ is well-defined. Variances ($\var  \theta$) were computed as the variance of the time-sequence $\{\theta(t_1),\theta(t_2),\ldots\}$. Whenever we refer to $\theta$ in the main manuscript without explicit reference to the time-dependence, we imply $\theta = \langle \theta (t) \rangle$ where we used a time average for $\langle \cdot \rangle$. Below we have reproduced a pseudocode to measure $\theta$:
	
	\begin{algorithm}
		\small{
			\SetKwInOut{Input}{Input}\SetKwInOut{Output}{output}
			\Input{ slip-link sites $\{ (i,j)_n \}, n=1,\ldots,N_p$ (with $N_p$ the number of slip-links). 
			}
			\Output{ loop network topology order parameter $\theta \in [0,1]$ }
			
			\BlankLine\BlankLine
			
			$n_\mathrm{nested} \leftarrow 0$\;
			
			\For{ $n\leftarrow 1$ \KwTo $N$ }{
				get slip-link sites $(i_n,j_n)$ of slip-link $n$\;
				\BlankLine
				compute self-loop-size of slip-link $n$: $\Delta_\mathrm{self} \leftarrow |i-j|$\;
				\BlankLine
				get sites $(i,j)_{n-1, n+1}$ of neighboring slip-links $n-1, n+1$\;
				\BlankLine
				compute nested-loop-size between slip-link $n$ and $n-1,n+1$: 
				\begin{align*}
					\Delta_{(n,n-1)} = & |i_n - i_{n-1}| + |j_n - j_{n-1}|\\
					\Delta_{(n,n+1)} = & |i_n - i_{n+1}| + |j_n - j_{n+1}|
				\end{align*}
				
				$\Delta_\mathrm{nested} \leftarrow \min(\Delta_{(n,n-1)} , \Delta_{(n,n+1)} )$ \;
				\BlankLine
				\If{$\Delta_\mathrm{nested} < \Delta_\mathrm{self}$}{
					$n_\mathrm{nested} \leftarrow n_\mathrm{nested} + 1$\;
				}
				
				\BlankLine

			}
			\BlankLine
			$\theta \leftarrow n_\mathrm{nested} / N_p $\;
			\BlankLine
			\Return $\theta$
			
			\BlankLine		\BlankLine		
			\caption{Pseudocode for computing the loop-network order parameter $\theta$. In this algorithm, we use the convention $|\Delta| = \min(\Delta, N_m - \Delta)$.}
		}	
	\end{algorithm}
	
	\subsection*{Determination of typical length of cross-diagonal}
	
\begin{figure}[t!]
		\begin{centering}
			\includegraphics[width=8cm]{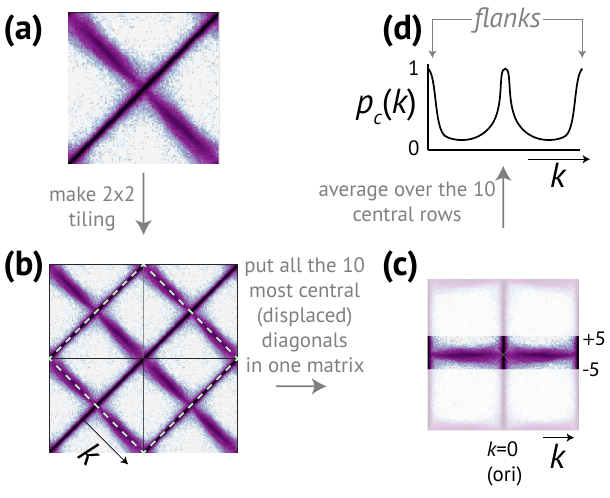}
			\par\end{centering}
		\caption{Illustration of our procedure to extract the cross diagonal from a
			contact map. \textbf{(a)}: A single contact map. \textbf{(b)}: Make a $2\times 2$ tiling of contact maps. \textbf{(c)}: Extract the 10 most central cross-diagonal rows (similar to the procedure published in \cite{Wang2017a, Le2013}). \textbf{(d)}: Average the data from step (c) over the 10 rows, returning a cross-diagonal probability $p_c(k)$ where $k=-\tfrac{1}{2}N_m,\ldots,+\tfrac{1}{2}N_m$ is the distance from \textit{ori}.
			In order to isolate the cross-diagonal from the rest
			of the contact map, we select the 10 most central rows from the contact map (i.e. $|k|\leq5$).
			\label{fig:getting-cross-diagonal-from-map}}
		
	\end{figure}

	We first extract an `unprocessed' cross-diagonal probability $p_{c, \mathrm{unprocessed}}(k)\in[0,1]$ as illustrated in Fig.\,S\ref{fig:getting-cross-diagonal-from-map}. Neither the moments $\langle k^n \rangle$ nor the $p-$th percentile of the `unprocessed' distribution  $p_{c, \mathrm{unprocessed}}(k)$  correlated well with the cross-diagonal width based on visual inspection. The reason for this is that $p_{c, \mathrm{unprocessed}}(k)$ contains large flanks at $|k|\approx N_m/2 \gg 1$ (Fig. S\ref{fig:getting-cross-diagonal-from-map}d) that were found to significantly impact estimates of $\hat X_c$.
	
\begin{figure}[b!]
		\begin{centering}
			\includegraphics[width=8cm]{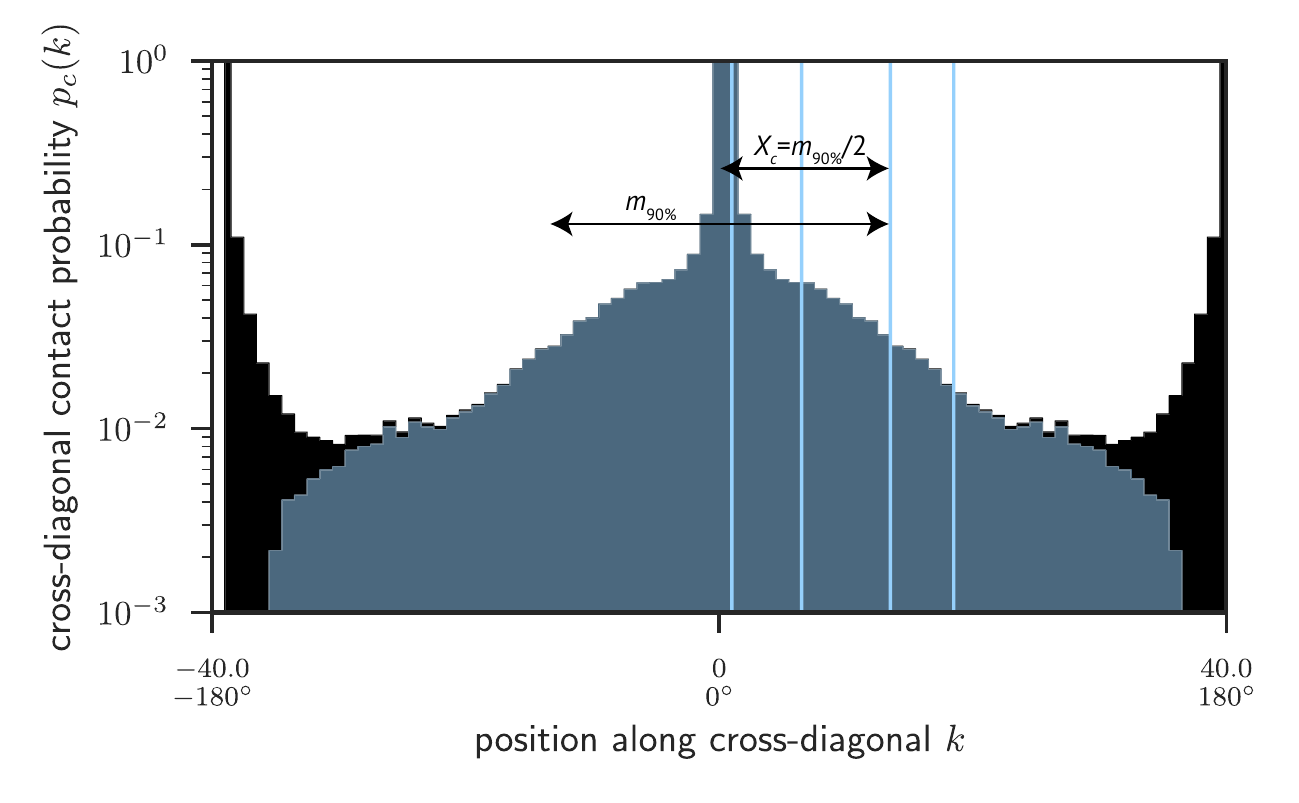}
			\par\end{centering}
		\caption{\textbf{Computation of the length of the cross-diagonal using the $p-$th percentile. }We first measured the `unprocessed' cross-diagonal contact probability $p_{c, \mathrm{unprocessed}}(k)$ from contact maps (black, see Fig.\,S\ref{fig:getting-cross-diagonal-from-map}
			for the procedure). Then, we subtract an estimate for the left and
			right flanks (blue, see section~\ref{sec:measurement-of-metrics-from-simulation-data}) and computed the $p-$th percentile $m_{p}$ of $p_{c}(k)$.
			The half-width of the $p-$th percentile $\tfrac{1}{2}m_{p}$ is shown for $p=50,75,90,95\%$
			(blue vertical lines, from left to right), and the typical width of
			the curve is defined as $X_{c}\equiv\tfrac{1}{2}m_{p}$. \label{fig:computing-width-of-cross-diagonal}}
	\end{figure}

	The flanks in $p_{c, \mathrm{unprocessed}}(k)$ (Fig. S\ref{fig:getting-cross-diagonal-from-map}d)  arise due to the circular topology of the polymer, with an increasing contact probability for $|k|\approx N_m/2 \gg 1$. A naive estimate of the flanks is a power-law with Flory-scaling $p_{c,\mathrm{naive}} \sim(\tfrac{1}{2}N_{m}-|k|)^{d\nu}, d=3,\nu \approx 3/5$ \cite{DeGennes1979}. Subtracting this naive estimate from  $p_{c, \mathrm{unprocessed}}(k)$,  was found to sometimes lead to negative probabilities. To avoid this, we instead subtracted an underestimate of the flanks, namely a power-law with $50\%$ stronger scaling than the naive estimate: $p_{c,\mathrm{underest.}}(k)\sim(\tfrac{1}{2}N_{m}-|k|)^{-1.5d\cdot\nu} \leq p_{c,\mathrm{naive}}$ (Fig. S\ref{fig:computing-width-of-cross-diagonal}, compare black and blue histograms). By visual inspection on a variety of representative test-cases, we found that this method indeed suppressed the flanks of $p_{c, \mathrm{unprocessed}}(k)$, but left the cross-diagonal itself intact (Fig. S\ref{fig:computing-width-of-cross-diagonal}, blue histogram).
	
	After subtracting the flanks from $p_c(k)$ (Fig.\,S\ref{fig:computing-width-of-cross-diagonal}, blue histogram), we calculate the typical length $X_{c,p}=\tfrac{1}{2}m_{p}$ 
	with parameter $p$ as the $p-$th percentile of the distribution $p_c(k)$. We computed various percentiles $\{m_{50\%},m_{75\%},m_{90\%},m_{95\%}\}$
	on a representative collection of contact maps and empirically found by visual inspection
	that $m_{75\%}$ was a good measure  for the
	typical length of the cross diagonal in contact maps. 
	
	\section{Universal scaling of slip-link propagation length for diffusive slip-links}\label{sec:universal-scaling-propagation-length}
	
\begin{figure}[t!]
		\begin{centering}
			\includegraphics[width=8cm]{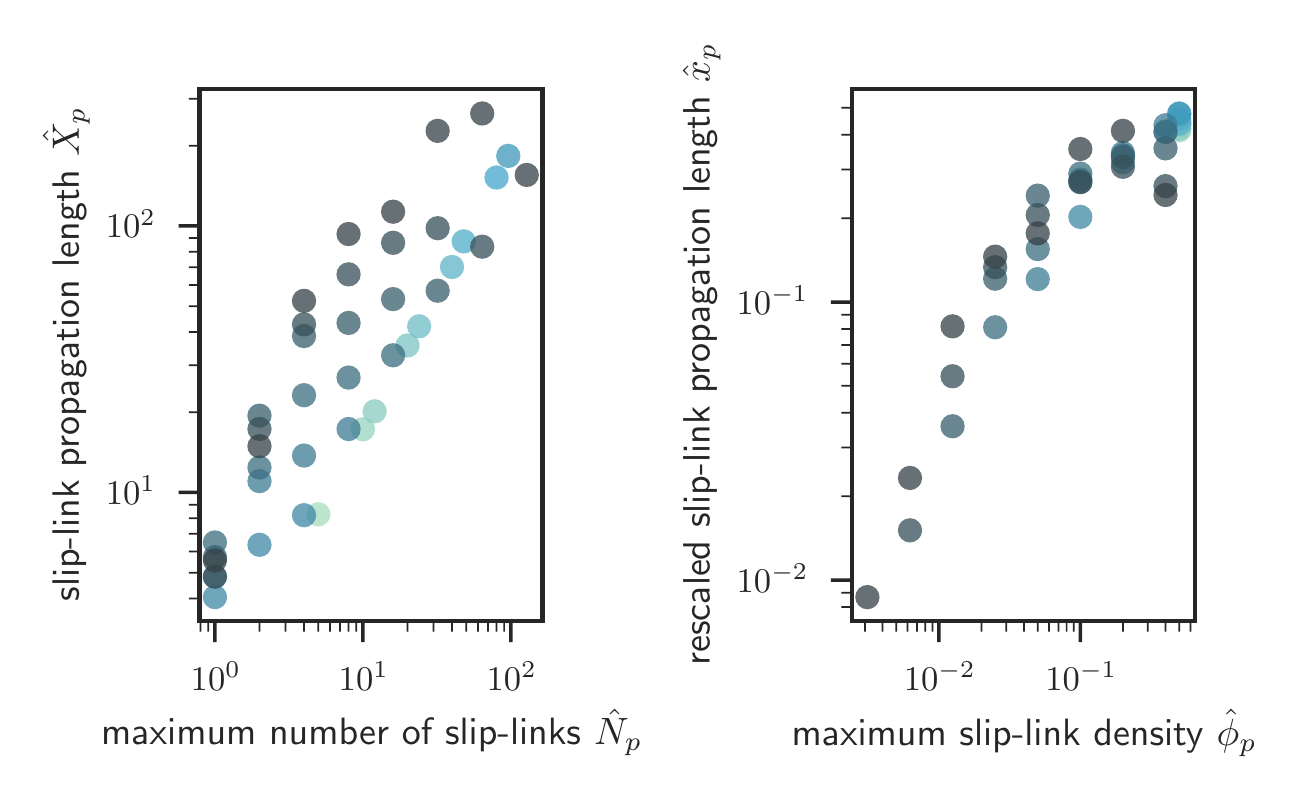}
			\par\end{centering}
		\caption{ \textbf{Left}: The data of $\hat X_p $ is not a single-valued function of $\hat N_p$, but also depends on $N_m$. \textbf{Right}: Rescaling both axes to intensive quantities $\hat x_p, \hat \phi_p$ (see text) collapses the data onto a single master curve. \label{fig:universal-scaling-of-xp_vs_phip}}
		
	\end{figure}

	The propagation length $\hat X_p $ and the number of slip-links $\hat N_p$ depend on $N_m$ (Fig. S\ref{fig:universal-scaling-of-xp_vs_phip}, left). For diffusive slip-links in the regime of fast dissociation kinetics, the relation between their scaled intensive counterparts $\hat x_p = \hat X_p / N_m, \hat \phi_p = 2N_p/N_m$  appear to be described by a single master curve (Fig.\,S\ref{fig:universal-scaling-of-xp_vs_phip}, right).
	
	\begin{figure}[t!]
		\centering
		\includegraphics[width=8cm]{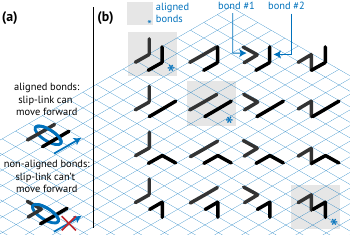}
		\caption{\textbf{Illustration of our simple counting argument for estimating $C$, the probability for two bonds to be aligned parallel.} \textbf{(a)}: A stiff slip-link can translocate in a certain direction (arrow) if the two polymer bonds in that direction are parallel. \textbf{(b)}: We count three possible configurations (marked~"*") for bonds \#1, \#2 that allow slip-link movement out of a possible 16 configurations. This implies that $C\approx 3/16$.}
		\label{fig:illustration-counting-argument}
	\end{figure}

	\section{Stability of juxtaposed organization by diffusive	slip-links}\label{section:relaxatin-juxtaposed-organization}
\begin{figure}[t!]
		\begin{centering}
			\includegraphics[width=8cm]{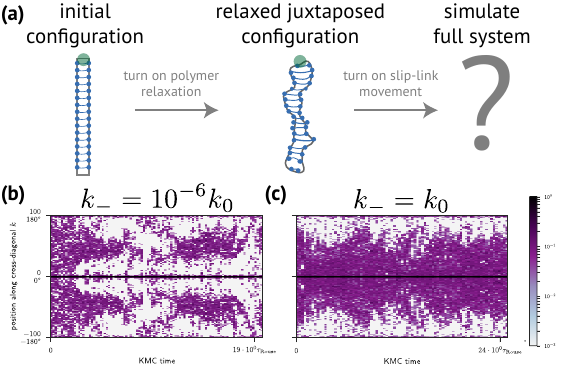}
			
			\caption{\textbf{The juxtaposed organization is only stable for fast dissociation kinetics.} \textbf{(a)}: Illustration of our simulation to text the stability of the juxtaposed organization. We first initialize a polymer in the juxtaposed organization with immobilized diffusive slip-links, but allow for polymer relaxation. We then turn on slip-link diffusion along the polymer, and recorded kymographs of the cross-diagonal from contact maps. \textbf{(b)}: Kymograph of the cross-diagonal for slow dissociation kinetics. The cross-diagonal disappears over time (see also SI movies 5[a--f]). \textbf{(c)}: Kymograph of the cross-diagonal for fast dissociation kinetics. The cross-diagonal persists (see also SI movies 4[a--f]). For both kymographs (b--c) we used $N_m=200, N_p=32$. \label{fig:illustration-relaxation-experiment} }
			
		\end{centering}
	\end{figure}

	To investigate the stability of the juxtaposed state with diffusive slip-links, we prepared a polymer in a juxtaposed organization. This was achieved by positioning immobilized slip-links at regular intervals along the polymer but nevertheless allowing for polymer relaxation (Fig. S\ref{fig:illustration-relaxation-experiment}a). After we turned on slip-link movement, we recorded kymographs of the cross-diagonal.

	Kymographs clearly reveal that the cross-diagonal in the regime of slow dissociation propagates away from the loading site (Fig.\,S\ref{fig:illustration-relaxation-experiment}b). The region around \textit{ori} in the contact maps slowly loses the anomalously high contact probabilities over time, meaning that the cross-diagonal disappears for diffusive slip-links in the slow dissociation regime. For slip-links in the fast dissociation regime, however, the cross-diagonal persists and remains stable over time for $N_p$ above a critical value (Fig.\,S\ref{fig:illustration-relaxation-experiment}c). Ensemble averaged movies of contact maps for $N_p=1,2,4,8,16,32$ confirm that the cross-diagonal only persists in the fast dissociation regime (SI movies 4[a-f] for fast dissociation and 5[a-f] for slow dissociation).

	\section{Cell anisotropy and confinement does not explain juxtaposed organization}\label{section:cross-diagonal-from-cell-asymmetry}
	
	To investigate the role of an anisotropic confinement on the cross-diagonal, we use an ellipsoidal harmonic potential,
	\[
	V_\mathrm{conf.}(x,y,z) = \dfrac{1}{2} \left( k_x x^2 + k_y y^2 + k_z z^2 \right),
	\]
	where the $k_{x,y,z}$ specify the 'spring-constants' of the confinement for monomers with positions $x,y,z$.  The ellipsoidal potential simulates the combined action of the cell membrane and crowding agents. We assume cylindrical symmetry \cite{Marbouty2015}, $ x^2 = y^2 = r^2 $ where $r$ is the radius of the cylindrical confinement. We use a length-to-width ratio of 4, slightly larger than the typical length-to-width ratio $\approx 3$ of many bacteria \cite{Moran2010a}.
	
\begin{figure}[t!]
		\includegraphics[width=8cm]{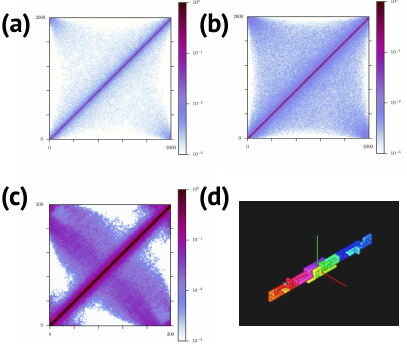}
		\caption{\textbf{Physiologically relevant confinement potentials do not result in a cross-diagonal.} \textbf{(a--b):} contact maps from polymers with $N_p=0$ (a) and $N_p=100$ (b) equilibrium slip-links with a confinement potential of widths $(\sqrt{  x^2 }, \sqrt{y^2 }, \sqrt{ z^2 })=(5\ell_0,5\ell_0,20\ell_0)$, where $\ell_0$ is the monomer length. Polymer length $N_m = 2000$. \textbf{(c--d):} Simulation with a confinement potential of anisotropy ratio $\sqrt{100}\approx 31.6$. Polymer length $N_m=200$. See SI movie 6 for a movie of the polymer for various other anisotropies. \label{fig:eq-simulations-effect-of-cell-anisotropy}  }
	\end{figure}

	We additionally include a spring-like tether capturing the effect of proteins, such as DivIVA in \textit{B. subtilis} and PopZ in \textit{C. crescentus}, that localize \textit{ori} to one of the cell poles \cite{Wang2014}. We use equilibrium Monte-Carlo simulations \cite{Broedersz2014} to investigate the organization of DNA in the presence of a confinement potential and tether. We find that the contact maps do not exhibit a cross-diagonal. The confinement potential does result in a slight asymmetry in the contact maps that increases weakly with the number of slip-links $N_p$ (Fig.\,S\ref{fig:eq-simulations-effect-of-cell-anisotropy}a--b). This slight asymmetry resembles a cross-diagonal under conditions  of large $N_p$ and small polymer-length $N_m$, but only becomes visually apparent in our contact maps when the density of slip-links is 10\%, $>100$ times higher than the SMC condensin density \textit{in vivo} \cite{Wilhelm2015}. Even so, even at this unphysiologically high concentration of slip-links, the asymmetry in the contact maps is very diffuse compared to that of the \textit{in vivo} Hi-C maps (Fig. S\ref{fig:eq-simulations-effect-of-cell-anisotropy}a--b). The width of the off-diagonal structure broadens as we increase $N_m$. Indeed, the cross-diagonal for $N_m=200$ is only clearly resolved when we increase the asymmetry of the confinement potential to unrealistically high values $\sqrt{z^2 /r^2 }  > 30$ (Fig.\,S\ref{fig:eq-simulations-effect-of-cell-anisotropy}c--d). In sum, our simulations indicate that ellipsoidal confinement and tethering alone cannot be responsible for a cross-diagonal.

	\section{Relationship between $\theta,\hat{X}_{p}$ and $\hat X_{c}$\label{sec:Relationship-between-Xc-Xp-and-theta}}
	
\begin{figure}[b!]
		\includegraphics[width=8cm]{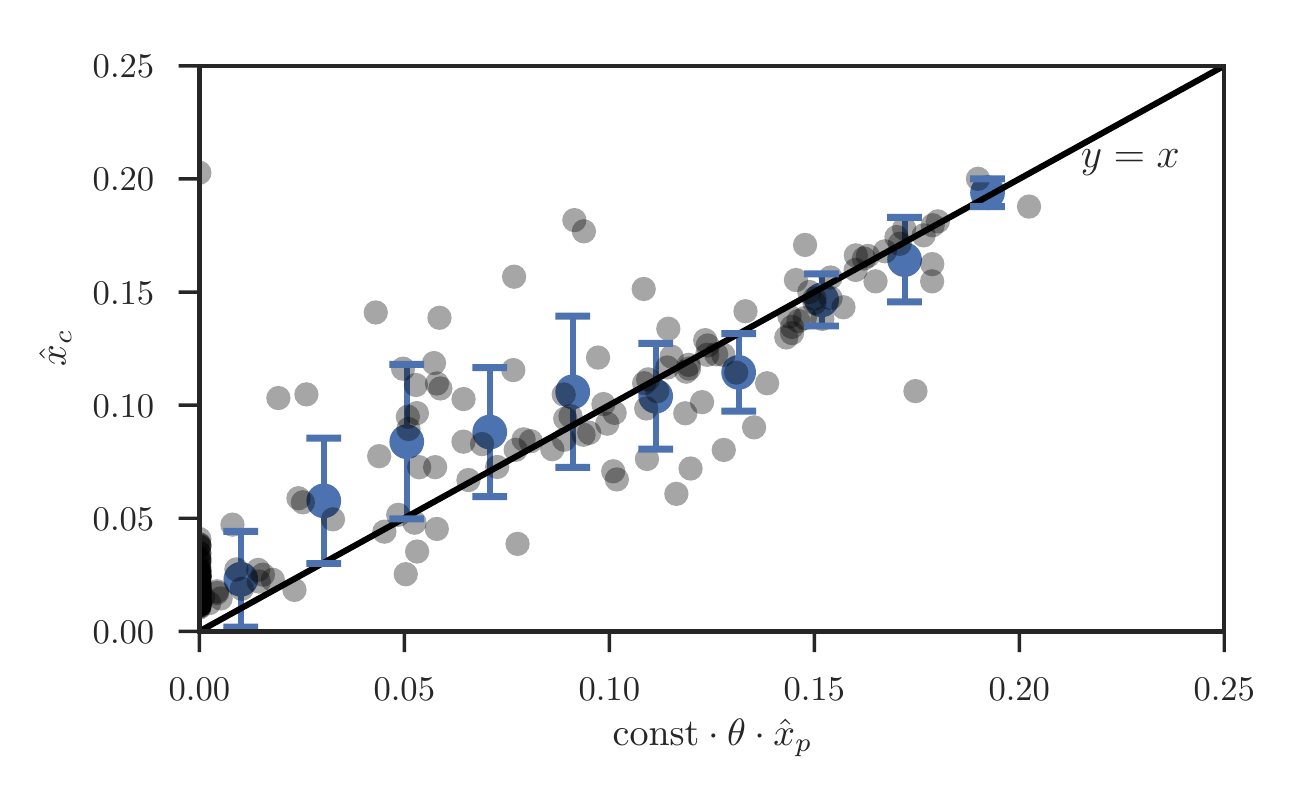}
		\caption{ The relative cross-diagonal width is $\approx \mathrm{const.} \theta \hat x_p$ with $\mathrm{const.}=1.3$. Error bars: standard deviation $\sqrt{ \mathrm{var} \hat x_c }$. \label{fig:xp-times-theta-vs-xc} }
	\end{figure}

	Our simulations reveal that the cross-diagonal
	length $\hat{x}_{c}$ is well approximated by an
	`and-gate' of $\theta$ and $\hat{x}_{p}$ (Fig. S\ref{fig:xp-times-theta-vs-xc}). This shows that one needs cooperative loops (high $\theta$) that also propagate deep
	into the bulk of the polymer (high $\hat x_p$).
	
	\section{Loop topology of polymers with diffusive slip-links}\label{sec:diff-slip-links-contact-distribution}

\begin{figure}[h!]
		\begin{centering}
			\includegraphics[width=8cm]{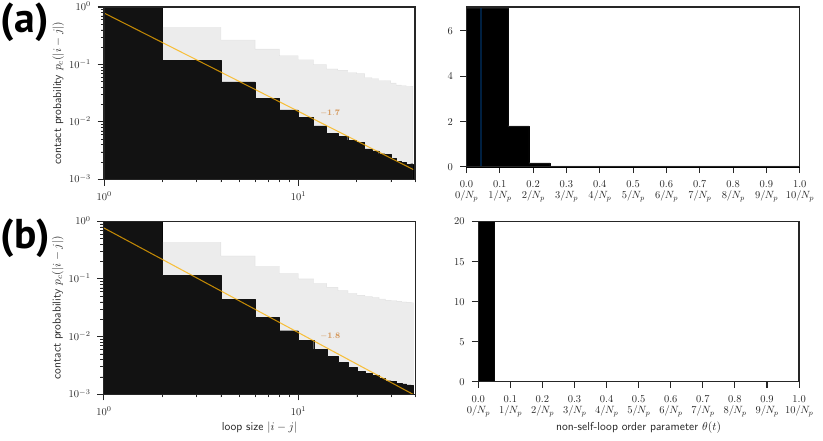}
			\caption{Left panels: Contact distribution $p(\Delta)$ for loop-size $\Delta$. Right panels: Distributions $p(\theta)$ of loop nesting parameter $\theta$. \textbf{(a):} Diffusive slip-links that bind non-specifically to the polymer (\dParBS). \textbf{(b):} Diffusive slip-links that bind exclusively to the loading site, but with slow dissociation kinetics ($k_- = 10^{-6}k_0$).
				\label{fig:contact-distribution} }
		\end{centering}
	\end{figure}

	We implemented a system with diffusive slip-links that can bind and unbind anywhere ($\Delta\mathrm{ParB}S$), so that this system obeys detailed balance and hence relaxes into thermodynamic equilibrium. The scaling  we observe for the \dParBS scenario $p \sim \Delta^{1.7}$ (Fig. S\ref{fig:contact-distribution}a, left) is nearly identical to that of a random polymer $p \sim \Delta^{d \nu}, d\nu \approx 1.8$ \cite{DeGennes1979}. The distribution of $\theta$ is highly peaked at $\theta=0$, showing that the diffusive slip-links almost always enclose self-loops and almost never enclose nested loops (Fig.\,S\ref{fig:contact-distribution}a, right). The system with diffusive slip-links that can only bind to the loading site breaks detailed-balance, but we find that for slow dissociation kinetics, the resulting distribution of loop-sizes $p\sim \Delta^{1.8}$ is nevertheless consistent with that of a random polymer (Fig.\,S\ref{fig:contact-distribution}b).
	
	\section{Diffusing cross-diagonal for motor slip-links without a loading site}\label{sec:motors-without-ori}

	Time-averaged contact maps of a polymer with motor slip-links but no loading site (\dParBS) are, apart from the main diagonal, homogeneous. Upon closer inspection, however, movies of the contact maps (SI movie 7a) suggest a diffusing cross-diagonal. Additionally, movies of the loop diagrams show that the motor slip-links extrude large loops until they collide and block further motion (SI movie 7b). To investigate this quantitatively, we perform Hough transforms of the contact maps. The Hough transform maps Cartesian space $(x,y)$ into Hough-space $(r,\theta)$. Lines in Cartesian can be represented as $r = x \cos \theta + y \sin \theta$, and the Hough transformed image is simply the image of these $(r,\theta)$. Thus, we can easily identify the lines in our contact maps.
	
\begin{figure}[h!]
		\includegraphics[width=8cm]{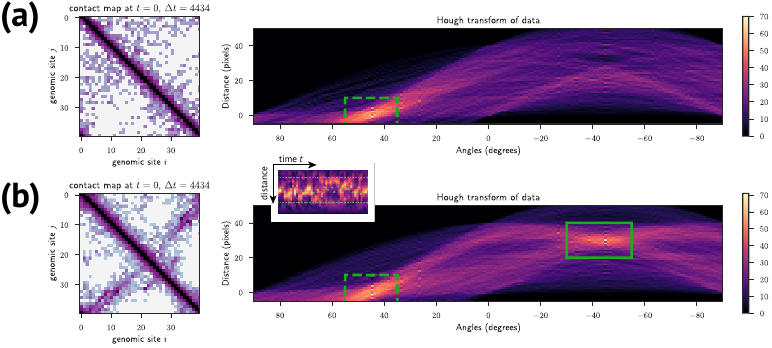}
		\caption{\textbf{Left panels}: Representative contact map at a time $t$ averaged over a time-interval $\Delta t$. \textbf{Right panels}: The Hough transform of the contact map. \textbf{(a):}~Simulation was performed with $\hat N_p=1$ with persistence time $\tau_\mathrm{switch}=10^{-3}$. \textbf{(b):}~Simulation was performed with $\hat N_p=2$ with persistence time $\tau_\mathrm{switch}=10^{-3}$. The Hough transforms in (b) displays an additional locus in the area $\theta=40^\circ\ldots 50^\circ, r  = 14 \ldots 42 \unit{px}$. This locus diffuses randomly over time (inset).\label{fig:hough-transform} }
	\end{figure}

	Intuitively, Hough transformed contact maps of this system with one motorized slip-link $\hat N_p = 1$ only contain a locus that corresponds to the main diagonal (Fig.\,S\ref{fig:hough-transform}a). Interestingly, however, when we have two motor slip-links $\hat N_p = 2$, there appears a very clear additional locus in the Hough transformed contact maps whose position diffuses randomly over time (Fig.\,S\ref{fig:hough-transform}b and SI movie 8).
	
\begin{figure}[h!]
		\includegraphics[width=8cm]{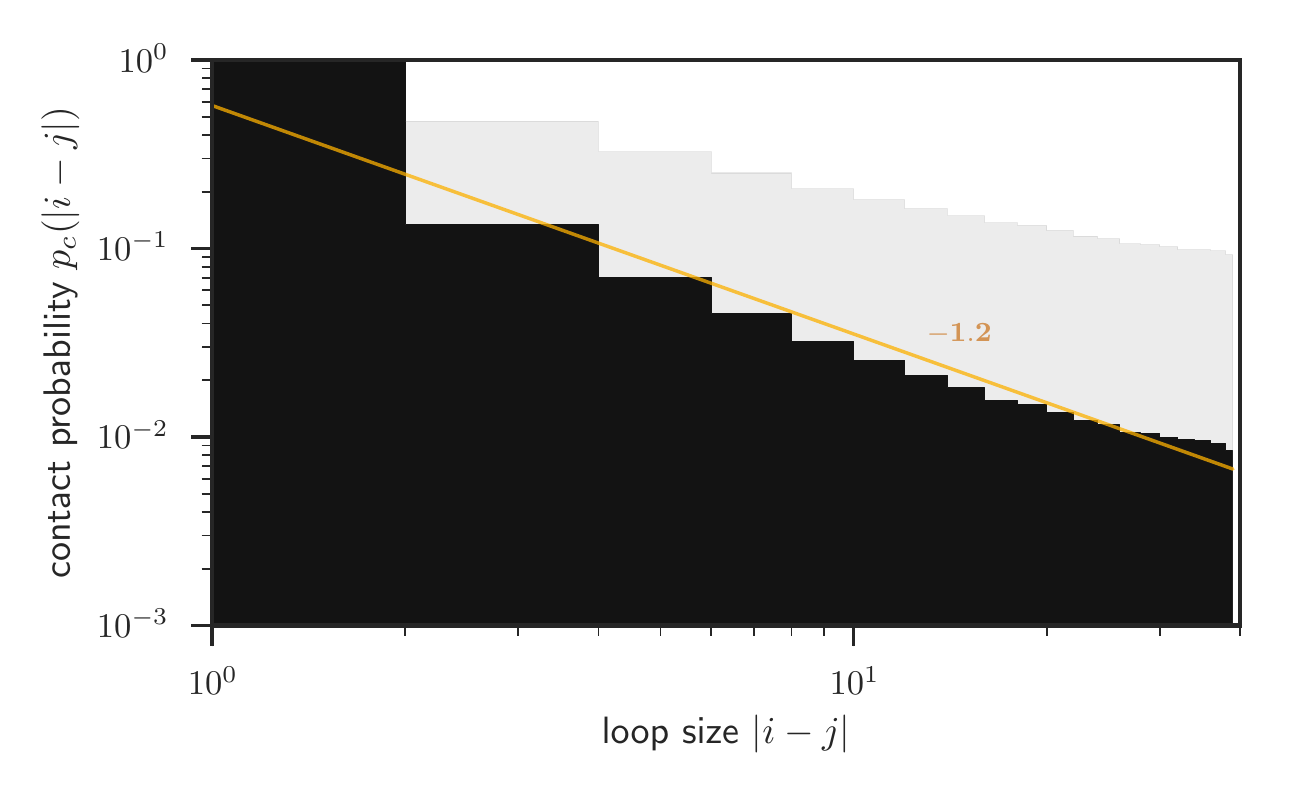}
		\caption{Contact distribution $p(\Delta)$ for loop-size $\Delta$ of a polymer with $N_p=10$ motor slip-links but lacking a loading site (\dParBS).\label{fig:contact-distribution-motor-wo-loading-site} }
	\end{figure}

The distribution of loop-sizes $p(\Delta)$ for the \dParBS system with motor slip-links appears to be characterized by an approximate power-law behavior for intermediate loop-sizes (Fig.\,S\ref{fig:contact-distribution-motor-wo-loading-site}). With the same number of slip-links, the exponent  of this powerlaw for the system with motor slip-links is smaller than that of the system with diffusive slip-links (compare Figs.\,S\ref{fig:contact-distribution} and S\ref{fig:contact-distribution-motor-wo-loading-site}). This change in the exponent associated to the power-law behavior of $p(\Delta)$ indicates that the \dParBS system with motor slip-links is characterized by a qualitatively different loop organization than that of a random polymer or that of a polymer organized by diffusive slip-links.
	
	\section{The impact of asymmetrical translocation on chromosome organization}\label{sec:asymmetric-loop-extrusion}
\begin{figure}[h!]
		\centering
		\includegraphics[width=6cm]{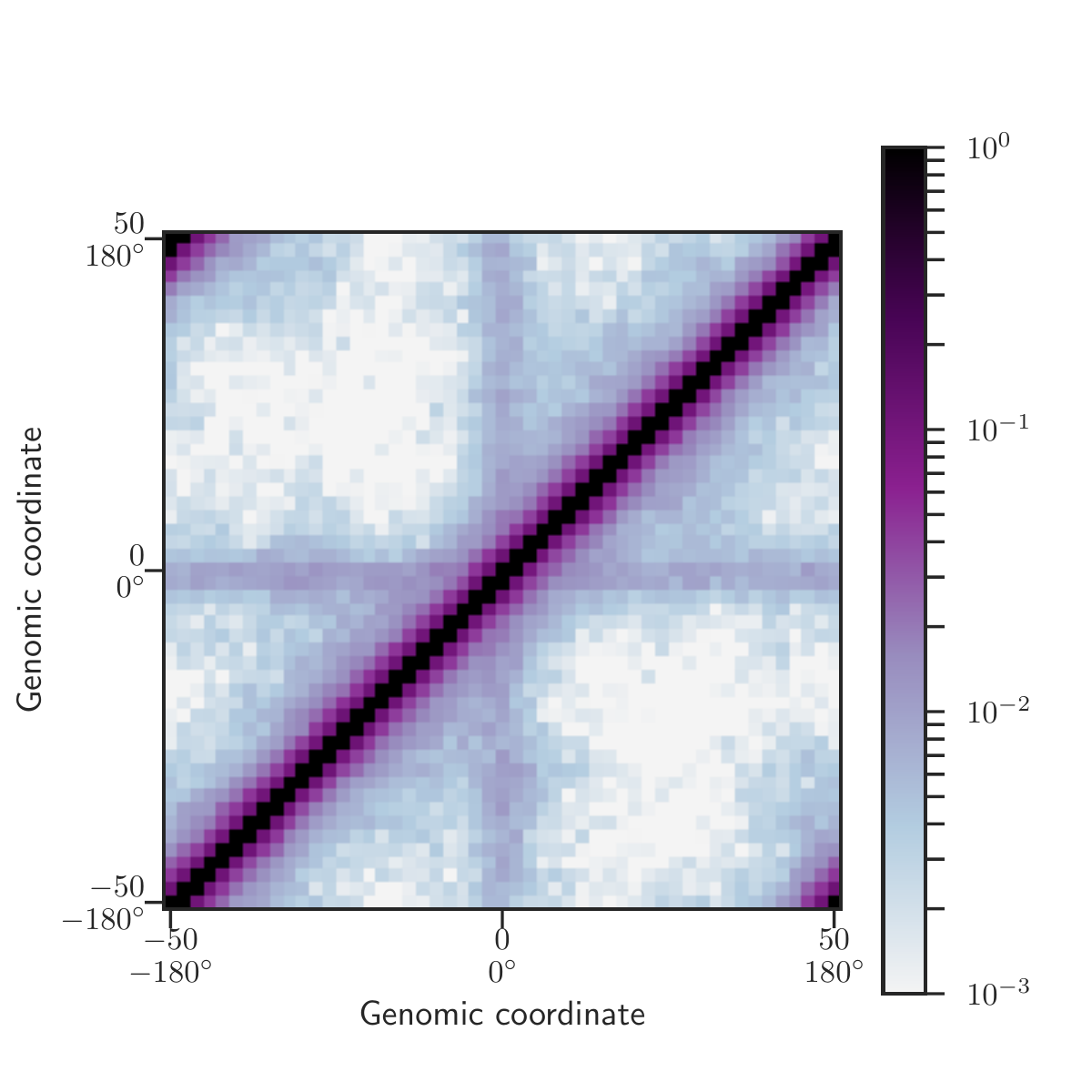}
		\caption{\textbf{Asymmetric loop extrusion results in a star-shaped pattern in contact maps.} For these data, we bound one asymmetric highly persistent motorized slip-link to the loading site. Note, in this simulation, none of the slip-links perform \textit{symmetric} loop extrusion, which is the type of motor activity discussed in the main text.}
		\label{fig:asymmetric-loop-extrusion}
	\end{figure}

	We implemented \textit{asymmetric} active translocation or loop extrusion as detailed below (see SI movie 9 for the dynamics of a loop diagram and the contact map in Fig.\,S\ref{fig:asymmetric-loop-extrusion}):
	\begin{enumerate}
		\item A slip-link is added to the loading site
		\item One of the two sides of the slip-link is immobilized
		\item The side of the slip-link that is not immobilized is assigned a random translocation direction (i.e. positive or negative).
	\end{enumerate}
	
	\bibliography{references}

	\newpage
	\section{Supplementary Movies}
All movies can be found at  \texttt{https://syncandshare.lrz.de/filestable/MlNmZTluQ2pDc0RkYWNhQ2NrQ3JH}

\includegraphics{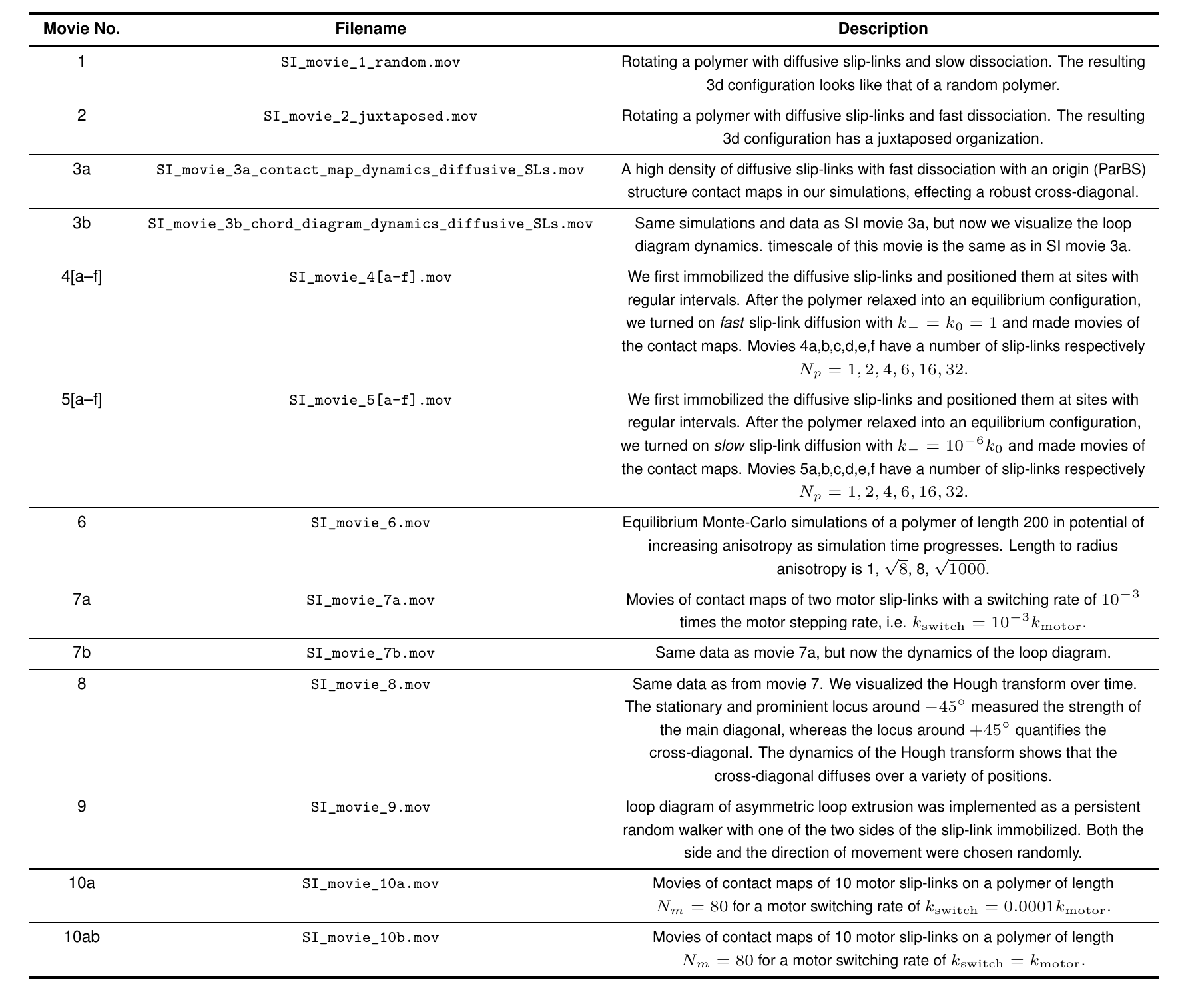}

\end{appendices}
	
\end{document}